%% file: paper.tex
\documentclass[letterpaper,twocolumn,10pt]{article}
\usepackage{usenix-2020-09}

\input{space-tricks}

\usepackage[utf8]{inputenc}
\usepackage[T1]{fontenc}
\usepackage[english]{babel}
\usepackage[binary-units=true]{siunitx}
\usepackage[ruled, vlined]{algorithm2e}
\usepackage{subcaption}
\usepackage{multirow}
\usepackage[inline]{enumitem}
\setlist{noitemsep,topsep=0pt,parsep=0pt,partopsep=0pt}
\usepackage{amsmath}

\usepackage{tikz}
\newcommand*\circled[2][]{\tikz[baseline=(char.base)]{
    \node[shape=circle,draw,inner sep=1pt,#1,every number/.try] (char) {#2};}}
\tikzstyle{every number}=[draw=blue] 

\newif\ifshowcomments
\showcommentstrue

\ifshowcomments
\newcommand{\mynote}[2]{\fbox{\bfseries\sffamily\footnotesize{\textbf{#1}}}
 {\small$\blacktriangleright$\textsf{\emph{#2}}$\blacktriangleleft$}}
\else
\newcommand{\mynote}[2]{}
\fi

\newcommand{\lkqq}{{Linux-KVM/Qemu/Qcow2}}
\newcommand{\lkq}{{Linux-KVM/Qemu}~}

\newcommand{\sys}{\textsc{sQemu}\xspace}
\newcommand{\vanilla}{\textsc{vQemu}\xspace}

\newcounter{numobserv}
\definecolor{beaublue}{rgb}{0.88, 0.93, 0.93}
\usepackage{tcolorbox}
\colorlet{shadecolor}{beaublue}
\tcbset{
     colback=beaublue,
     boxrule=0.5pt,
     boxsep=1pt,
     left=1pt,
     right=1pt,
     top=1pt,
     bottom=1pt,
     sharp corners,
     colframe=white,
 }
 \newcommand{\observ}[1]{
 \addtocounter{numobserv}{1}
 \begin{tcolorbox}
	\textit{\textbf{Take-away\,\thenumobserv:} #1 }
\end{tcolorbox}
}

\begin{document}

\title{Virtual Disk Snapshot Management at Scale}

\author{
	{\rm Kevin Nguetchouang Ngongang}\\
	ENS Lyon
	\and
	{\rm Stella Bitchebe}\\
	University of Nice
	\and
	{\rm Theophile Dubuc}\\
	ENS Lyon/Outscale 3DS
	\and
	{\rm Mar CALLAU-Zori}\\
	Outscale 3DS
	\and
	{\rm Christophe Hubert}\\
	Outscale 3DS
	\and
	{\rm Pierre Olivier}\\
	University of Manchester
	\and
	{\rm Alain Tchana}\\
	ENS Lyon
} 

\maketitle

\input{00-abstract}

\input{01-introduction}
\input{02-background}
\input{03-characterization}
\input{04-contributions}

\input{05-evaluation}

\input{06-related-works}

\input{07-conclusion}

\bibliographystyle{plain}
\bibliography{paper}

\end{document}

%% file: space-tricks.tex
\usepackage[compact]{titlesec}

\usepackage[subtle]{savetrees}

\usepackage{setspace}
\usepackage[margin=0pt,skip=7pt,font={stretch=0.9}]{caption}


%% file: 00-abstract.tex
\begin{abstract}
Contrary to the other resources such as CPU, memory and network, for which virtualization is efficiently achieved through direct access, disk virtualization is peculiar.
In this paper we make four contributions.
Our first contribution is the characterization of disk utilization in a public large scale cloud infrastructure.
It reveals the presence of long snapshot chains, composed sometimes of up to 1000 files.
Our second contribution is to show by experimental measurements that long chains lead to performance and memory footprint scalability issues.
Our third contribution is the extension of both the Qcow2 format and its driver in Qemu to address the identified scalability challenges.
Our fourth contribution is the thorough evaluation of our prototype, called \sys, demonstrating that it brings significant performance enhancements and memory footprint reduction.
For example, it improves the throughput of RocksDB by about 48\% compared to vanilla Qemu on a snapshot chain of length 500.
The memory overhead on that chain is also reduced by 15x.
\end{abstract}

%% file: 01-introduction.tex


\section{Introduction}
\label{introduction}
\begin{figure}
    \centering
    \includegraphics[width=.8\columnwidth]{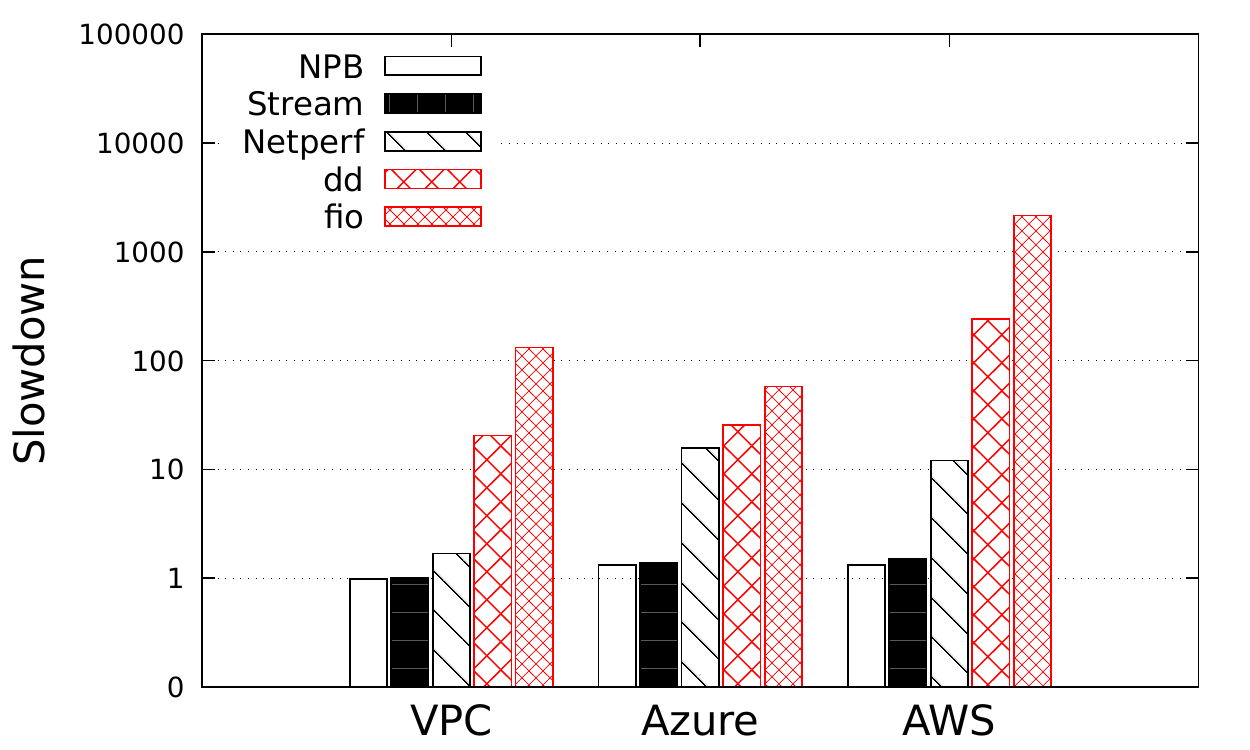}
    \caption{Performance slowdown incurred by virtualization for different types of applications.
    The results are presented in logarithmic scale. \textbf{Lower is better.}}
    \label{fig:perfomance-slowdown}
\end{figure}

Virtualization is the keystone technology making cloud computing possible and therefore enabling its success.
However, virtualization, and thus cloud computing, comes at the cost of a certain overhead on application performance.
That overhead have been well studied~\cite{XEN, VIRT_OVERHEAD1, VIRT_OVERHEAD2, VIRT_OVERHEAD3, VIRT_OVERHEAD4, VIRT_OVERHEAD5, VIRT_OVERHEAD6, VIRT_OVERHEAD7}.
Although it concerns all types of resources (CPU, RAM, network, disk), they are not all affected with the same intensity.
Figure~\ref{fig:perfomance-slowdown} shows the performance degradation coming from virtualization for a wide range of benchmarks including Stream~\cite{STREAM} (memory intensive), NPB~\cite{SNU_NPB} (CPU intensive), netperf~\cite{NETPERF} (network intensive), as well as the Linux \texttt{dd} command (disk intensive, throughput-oriented) and fio~\cite{FIO} (disk intensive, latency-oriented), when they run in AWS EC2 (\texttt{t2.medium} instance type), Microsoft Azure (\texttt{Standard\_B2s} instance type), a virtualized private cloud, and on bare metal private cloud without virtualization\footnote{We chose \texttt{t2.medium} and \texttt{Standard\_B2s} to match the VM size that we used in our private cloud.}.
We use the latter as the baseline.
We can observe that the two disk-intensive applications (\texttt{dd} and \texttt{fio}) experience the highest slowdown.
For \texttt{fio}, it is about 1,639$\times$ the degradation experienced by NPB.


Surprisingly, contrary to the other resource types, very few research work focuses on improving storage virtualization in the cloud.
It is important to fill this gap, in particular in the context of the explosion of data-centric application's (big data, ML and AI trends) popularity.
Disk virtualization is peculiar as it is still implemented through complex multi-layered architectures~\cite{10.5555/2208461.2208469,10.1145/2597652.2597667}.
Another illustration of the singularity of disk virtualization is the fact that it is generally achieved through the use of complex virtual disk formats (Qcow2, QED, FVD, VDI, VMDK, VHD, EBS, etc.) that not only perform the task of multiplexing the physical disk, but also need to support standard features such as snapshots/rollbacks, compression, and encryption.
These indirections are the source of the disk virtualization overheads.

This paper focuses on \lkq~(hereafter LKQ), a very popular virtualization stack.
LKQ supports several virtual disk formats, among which Qcow2~\cite{QCOW2} is widely adopted in production~\cite{RAVELLO}.
Our cloud partner, which is a large scale public cloud provider with several datacenters spread over the world, relies on LKQ and Qcow2.
A salient feature provided by Qcow2 is the capacity to create incremental Copy-On-Write (COW) snapshots (backing files) in order to save the state of the virtual disk at a given point in time and to reduce storage space usage.
The virtual disk of a VM can thus be seen as a chain linking multiple backing files.
In this paper, we identify and solve virtualization scalability issues on such snapshot chains.

Our \textit{first contribution} (\S \ref{scalability-issue}) is the characterization of disk usage in the infrastructure of our cloud partner.
We found that snapshot operations are very frequent in the cloud (some VMs are subject to more than one snapshot creation per day) for three main reasons.
First, cloud users leverage snapshots to periodically create recovery points for fault tolerance reasons.
Second, cloud users and providers use snapshots to achieve efficient virtual disk copy operations, as well as to share some elements such as the OS/distribution base image between several distinct virtual disks.
Third, cloud providers use the snapshot feature to transparently distribute a virtual disk, made of multiple chained backing files, among several storage servers, in effect going above the boundaries of a single physical server.
This is achieved for load balancing reasons, and also tackles resource fragmentation issues.
For all these reasons, we observed that the length of a chain can be very high.
We identified chains composed of up to $1,000$ backing files.
To our knowledge, this is the first paper performing such characterization.
Prior works~\cite{ressource-central,azure-functions,harvested-vm} mainly focused on the characterization of virtualized CPU and memory utilization in the cloud.

Our \textit{second contribution} (\S \ref{sec:problem}) is to show by experimental measurements that long chains pose both performance and memory footprint scalability issues.
For illustration, using a synthetic benchmark based on \texttt{dd}, we measured up to 91\% of IO throughput decrease and up to 180$\times$ memory footprint increase for a chain composed of $1,000$ backing files.
We found that the origin of this problem lies in the fundamental design of the Qcow2 format:
the fact that the Qcow2 driver in Qemu manages each backing file \textit{individually in a recursive fashion}, without a global view of the entire chain composing the virtual disk.
The analysis of other formats (such as FVD~\cite{FVD}, see \S \ref{sec:rw}) shows that they use a similar approach.


Our \textit{third contribution} (\S \ref{design}) is to address these scalability by evolving Qcow2 and introducing two key principles:
1) direct access upon an I/O request, regardless of their position in the chain;
2) the use of a single Qcow2 metadata cache, avoiding memory duplication by being independent of the chain length.
The implementation of these principles raises three challenges.
First, we should allow backward compatibility, which is necessary to facilitate the adoption of our solution by cloud operators.
Second, we should preserve all Qcow2 features including compression, encryption, etc.
Third, important optimizations such as prefetching that come naturally with the current Qcow2 format should also be preserved.
To cope with the above challenges, we slightly extend the Qcow2 format in order to indicate, for each cluster of the virtual disk, the backing file it is contained in.
To preserve backward compatibility, we rely on reserved bits in Qcow2's metadata.
We implement these principles by extending on the one hand the Qemu's Qcow2 driver and the snapshot operation on the other hand.
We thoroughly evaluate our prototype in several situations: various disk sizes, chain lengths, cache sizes, and benchmarks.
Our solution tackles Qcow2's scalability issues regarding IO performance and memory footprint.
For example, on a virtual disk backed up by a chain of 500 snapshots, RocksDB's throughput is increased by 48\% versus vanilla Qemu.
The memory overhead on that chain is also reduced by 15$\times$.

Overall we make the following contributions:
\begin{itemize}
    \item We characterize for the first time virtual disk management and usage in a large scale could provider.
    \item We assess for the first time performance/memory footprint scalability issues in \lkqq, a popular virtualization stack, and explain the origin of the problems.
    \item We introduce two principles for addressing this problem.
    \item We implement these principles in Qemu while preserving all its features.
    \item We evaluate our prototype in various situations, demonstrating the effectiveness of our approach.
\end{itemize}

The rest of the paper is organized as follows.
\S \ref{background} presents the background.
\S \ref{scalability-issue} presents virtual disk characterization results.
\S \ref{sec:problem} presents and assesses the scalability issues handled in this paper.
\S \ref{design} presents our design to addresses the identified scalability issues.
\S \ref{evaluations} presents the evaluation results of our design.
\S \ref{sec:rw} presents the related work.
\S \ref{conclusion} concludes the paper.

%% file: 02-background.tex

\section{\lkqq}
\label{background}
Our goal is twofold:
the characterization of virtual disk management in a public large scale cloud and
handling of two scalability issues happening in long snapshot chains.
This section presents the necessary background to understand our contributions.

\begin{figure}
    \center
    \includegraphics[width=0.45\textwidth]{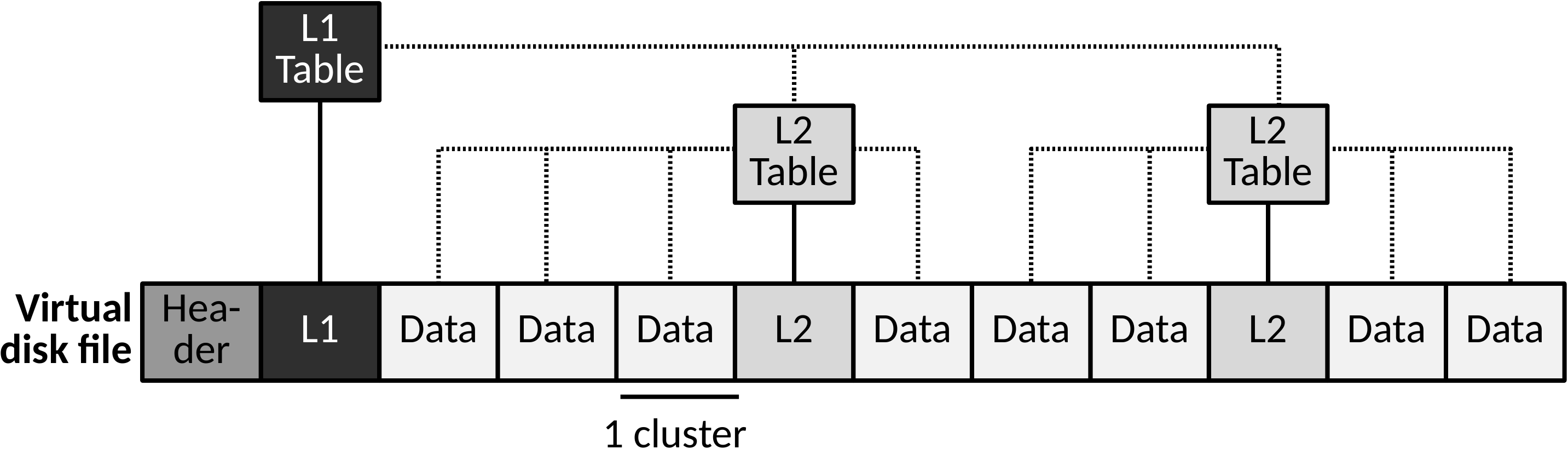}
    \caption{Overview of the Qcow2 format.}
    \label{fig:qcow2-format}
\end{figure}
\begin{figure*}
    \center
    \includegraphics[width=.8\textwidth]{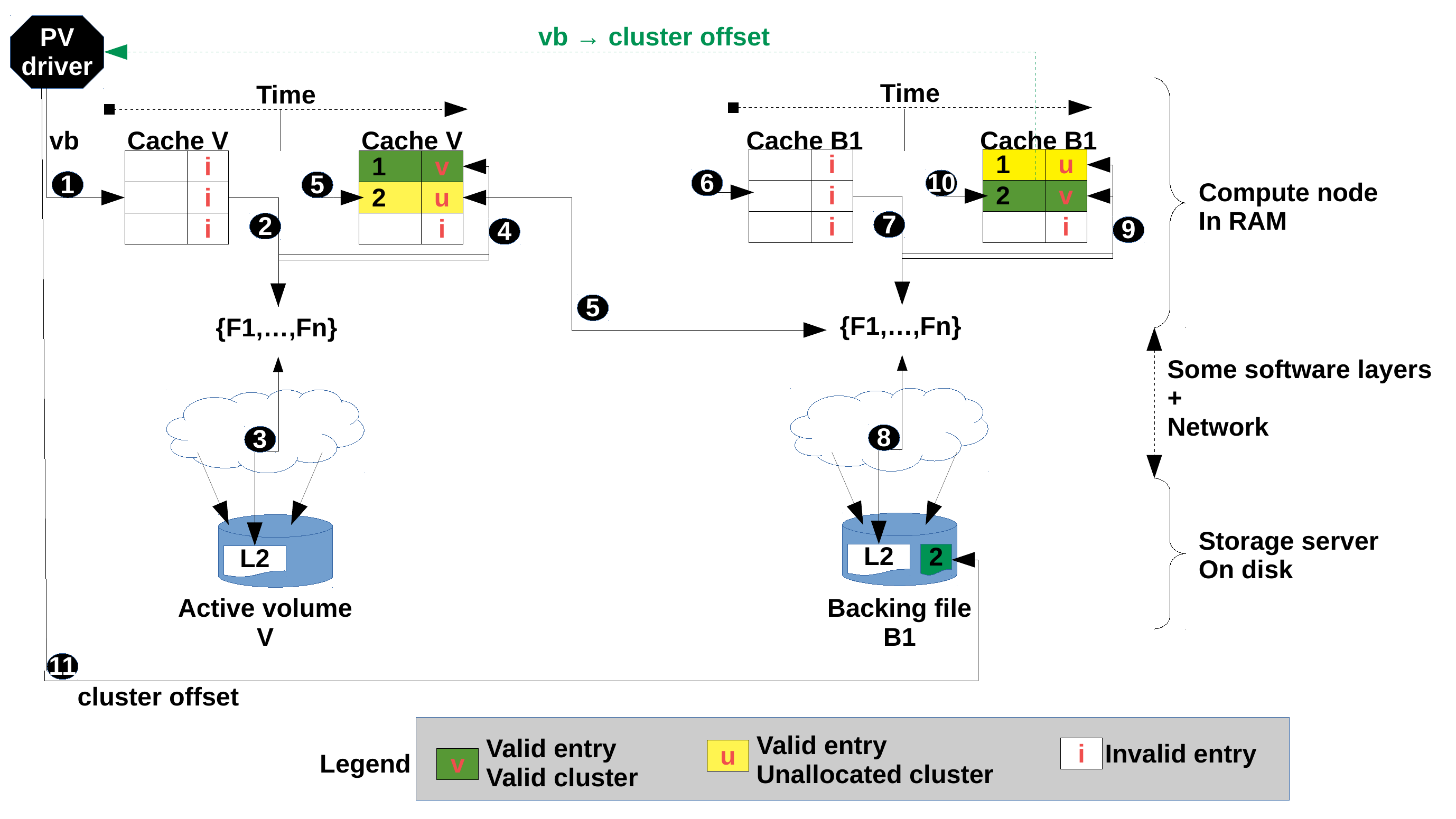}
    \caption{The journey of an IO request.}
    \label{fig:cache-walk}
\end{figure*}
\paragraph{Qcow2 Overview.}
The Qcow2 format enables copy-on-write snapshots by using an indexing mechanism implemented in the format and managed at runtime in the Qcow2 driver, running in Qemu, to map guest IO requests addressing virtual sectors/blocks to host offsets in the Qcow2 file(s).
An overview of the Qcow2 format is given on Figure~\ref{fig:qcow2-format}.
Without any snapshot, a virtual disk is contained in a single file.
The file is divided into units named clusters, that can contain either metadata (e.g, a header, indexation tables, etc.) or data that represent ranges of consecutive sectors.
The default cluster size is 64 KB.
Indexation is made through a 2-level table, organized as a radix tree:
the first-level table (L1) is small and contiguous in the file, while the second-level table (L2) could be spread among multiple non-contiguous clusters.
The header occupies cluster 0 at offset 0 in the file and the L1 tables comes right after the header.
For performance reasons, L1 and L2 entries are cached in RAM (see bellow).



\paragraph{Qcow2 Snapshotting.}
A Qcow2 virtual disk file $F$ can be linked to a backing file, i.e. a file that will be queried for clusters that are not present in $F$.
Today, the most common way to create a live incremental snapshot of a virtual disk $F$ for a given VM is to create a new empty Qcow2 file $E$ and set it as the current disk (called \textit{active volume}) for the VM while the previous virtual disk $F$ is set as the backing file for $E$.
In that way, all write operations made by the VM will be directed to the active volume ($E$) while read operations will be directed either to $E$ if the addressed sectors are present there, or to backing files if not.
With time, backing file chains can become very long (see \S \ref{scalability-issue}).


\paragraph{Qcow2 Cache Organization.}
To speed up access to L1 and L2 tables, Qemu caches them in RAM.
It creates and manages one cache for the active volume and one cache per backing file.
Each cache is managed independently from the others.
In the following, we describe how the cache works.
Qemu maintains a separate cache for the L1 table and a cache for L2 tables entries.
With its small size, the entire content of L1 is loaded in RAM at VM boot time.
The cache of L2 entries is populated on-demand, with a prefetching policy.
We therefore focus on the caching of L2 entries as they are likely to suffer from misses, thus influence IO performance.
When there is a cache miss, Qemu brings into the cache a set of L2 entries, a \emph{slice} of configurable size, among which the entry at the origin of the miss.
The slice is also the granularity of the cache eviction policy, which is LRU.
A cache entry includes: the file offset of the slice (noted \texttt{l2\_slice\_offset}), the number of threads which currently uses the slice (noted \texttt{ref}), the actual L2 entries composing the slice, and a field indicating whether a data cluster referenced by a L2 entry has been modified (noted \texttt{dirty}),
\texttt{l2\_slice\_offset} services as the tag when searching an entry in the cache, which is fully associative.


\paragraph{Qcow2 Cache Utilization.}
Every IO request issued by the guest OS to virtual disk $vb$ traps inside Qemu.
It is then handled by a thread running the para-virtualized disk driver in Qemu.
One of its main goals is to translate $vb$ to a data cluster offset inside the active volume or a backing file.

From $vb$, Qemu computes \texttt{l2\_slice\_offset}, \texttt{l2\_slice\_index}, and \texttt{l2\_index}.
Having this, Qemu looks if there is an entry in the cache that matches \texttt{l2\_slice\_offset}.
If it exists, then Qemu increments the corresponding \texttt{ref}.
Next, thanks to \texttt{l2\_slice\_index}, it reads the L2 entry.
If the latter describes an allocated data cluster (hereafter ``cache hit''), then Qemu reads the offset of the data cluster.
If the cluster is not allocated (hereafter ``cache hit unallocated'') then Qemu considers the cache of the next backing file in the chain.
If the slice is not in that cache, then Qemu will try to fetch it from the actual backing file associated with the current cache.
If the slice does exist on disk, then it is brought into the cache.
Otherwise, Qemu considers the cache of the next backing file, and so forth.

Some additional actions are performed for write requests.
First of all, the \texttt{dirty} field of the slice is set to 1.
If the L2 entry is found in a backing file (not the active volume), Qemu allocates a data cluster on the active volume and performs the copy-on-write.
If, despite the whole chain scanning, the L2 entry is not found, then Qemu just creates a new data cluster in the active volume.
In any case, Qemu configures L1 and L2 tables accordingly, both on disk and in the active volume's cache.
A cache entry can be evicted either when the VM is terminated or when the cache is full.

\paragraph{IO Request Journey on a Chain.}
Qemu manages a chain snapshot-by-snapshot, starting from the active volume.
Figure~\ref{fig:cache-walk} illustrates the journey of an IO request for a chain of size 2: the base image (B) and the active volume (V).
We assume that all L2 indexing caches are empty.
Let us assume that cluster number 2 is the target cluster, and it resides in B (meaning that it has not been modified since the creation of V).
\circled[fill=black,text=white]{1} The driver starts by parsing V indexing cache.
To handle the cache miss, Qemu performs a set of function calls with some of them \circled[fill=black,text=white]{3} accessing over the network the Qcow2 file to fetch the missed entry from V's L2 table.
According to its prefetching feature, Qemu fetches a slice of L2 table entries from V and \circled[fill=black,text=white]{4} fills V indexing cache.
In Figure~\ref{fig:cache-walk}, we assume that the size of a slice is 2 entries.
Thus, V's cache includes at the end of the first cache miss handling process two valid entries: cluster 1 and cluster 2.
After this step, \circled[fill=black,text=white]{5} pv driver hits V's cache, but the state of cluster 2 is marked unallocated because the references data cluster resides on B.
\circled[fill=black,text=white]{6} This cache hit unallocated event triggers the same Qemu functions used for handling a cache miss.
For cache hit unallocated events, Qemu moves to the parent snapshot (B).
In fact, at VM startup, Qemu initializes a linked list corresponding to the snapshot chain of the VM's virtual disk.
The caches of all the snapshots are also initialized at that time.
\circled[fill=black,text=white]{6}The first access to B's cache generates a miss \circled[fill=black,text=white]{7}.
After handling this miss (\circled[fill=black,text=white]{8}-\circled[fill=black,text=white]{10}), the offset of cluster 2 is returned to the driver.
From there, \circled[fill=black,text=white]{11} the latter can issue the IO request (consider read here).

%% file: 03-characterization.tex
\section{Virtual Disk Management Characterization}
\label{scalability-issue}
This section presents a characterization of virtual disk management by our cloud industry partner.
We explore various metrics, namely VM requested storage size, snapshot chain length, snapshot chain sharing, and snapshot creation frequency.
The study targets a datacenter located in Europe.
The number of VMs booted in 2020 in this region is 2.8 millions which corresponds to one VM booted every 12 seconds, demonstrating the large scale of our study.
The software and workloads running in the VMs are highly varied.
However it is worth noting that our partner specializes in business-to-business, hence the VMs run enterprise workloads as opposed to private individual ones.
Similar to existing cloud providers, the region runs VMs internal to our partner in addition to client VMs.
In the region, the VMs' virtual disks are backed up by Qcow2 chains.

\paragraph{Virtual Storage Resources Requested.}
\label{disk-usage}

\begin{figure}
    \center
    \includegraphics[width=0.45\textwidth]{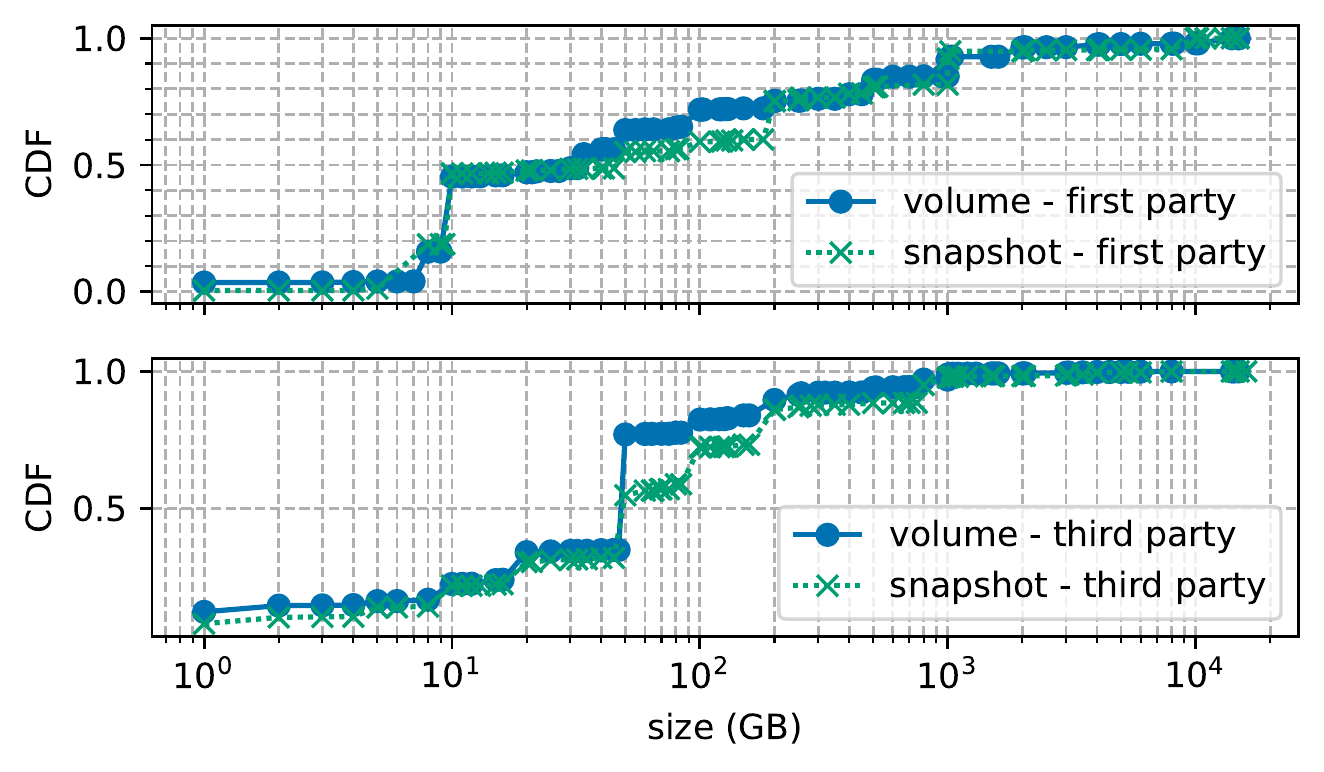}
    \caption{CDF of the virtual disk size for active volumes (\emph{volume}) and backing files (\emph{snapshots}) for first-party (provider) and third-party (client) VMs.}
    \label{fig:resources-requested-cdf}
\end{figure}
Figure~\ref{fig:resources-requested-cdf} shows the CDF of the virtual disk size for active volumes as well as backing files created over a day in 2020.
We have separated first party resources, used by the provider to operate the cloud and to provide other services, as well as third party resources, used by final customers.
We can see volumes and snapshots requesting up to 10 TB.
10 GB volumes corresponds to the default virtual disk size, and represents 30\% of the first party requests in both volumes and snapshots.
In the case of the third party, the most popular size among clients is around 50 GB with 40\%.

\observ{Virtual disks of all sizes, up to 10 TB, are used in the infrastructure. The most popular requested sizes are 10 GB (first party) and 50 GB (third party).}

\paragraph{Chain Length.}
\label{length}
Over the entire year 2020, we performed a daily measurement of the length of each chain in the infrastructure.
This covers a wide dataset, with the number of daily chains considered being in the order of the hundreds of thousands.
Snapshots (backing files) can be created by clients but also by the cloud provider, for example when it is decided that a new storage node should host part of a virtual disk, or when the client creates a virtual disk as a copy of another one.

\begin{figure}
    \center
    \includegraphics[width=0.45\textwidth]{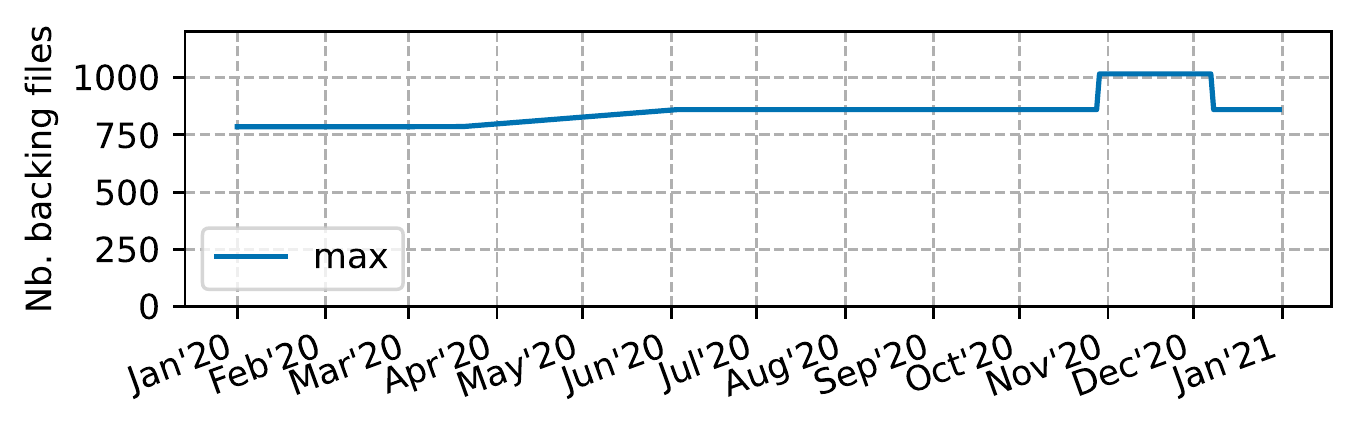}
    \caption{Evolution of the longest chain's size over the year 2020.}
    \label{fig:longer-chain-len}
\end{figure}

Figure~\ref{fig:longer-chain-len} presents the evolution of the longest chain's length over the period.
As we can see, there is always a chain with at least a length of 800 snapshots, and the longest chain can have a length of up to more than 1,000.

\begin{figure}
    \center
    \includegraphics[width=0.45\textwidth]{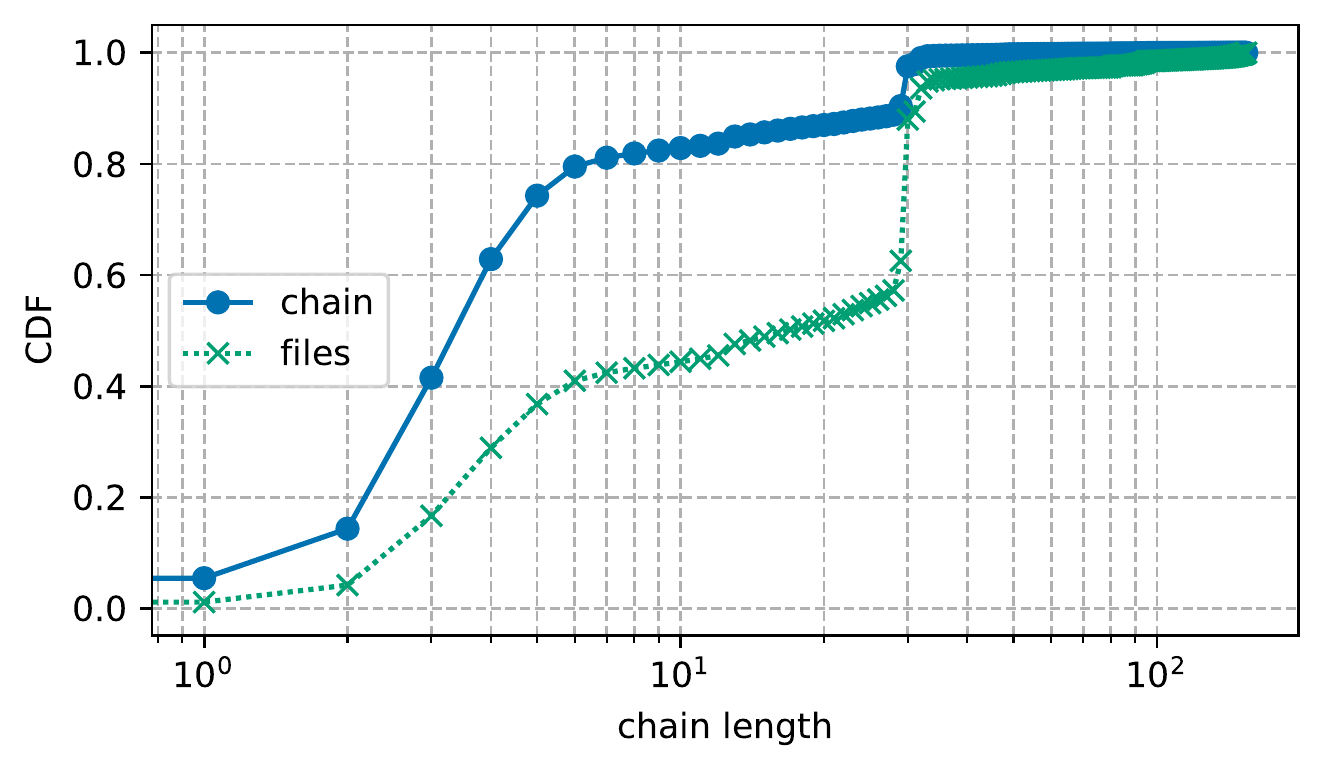}
    \caption{CDF of the chain length for a daily measurement.}
    \label{fig:chain-len-cdf}
\end{figure}

We studied in details a daily measurement made during the period when the longest chain was of a length superior to 1000.
Figure~\ref{fig:chain-len-cdf} shows the CDF for chains and files (active volumes and backing files) with respect to the chain length (for a file, to the length of the chain it belongs to).
Most of the chains are relatively small: chains of length 10 or lower represent nearly 50\% of the total number of files, and more than 80\% of the chains, in the platform.
A jump can be observed around size 30, with chains of size 30-35 files representing a relatively large proportion: 10\% of the chains and 25\% of the files.
This is because, for a subset of the chains, the backing file merging operation, named streaming, is triggered around size 30.
That operation merges the layers corresponding to multiple backing files into a single one.
The files that can be merged in this way correspond to unneeded snapshots, i.e. deleted client snapshots as well as the ones made by the provider.
Streaming helps reduce the size of some chains, however note that valid (non-deleted) client snapshots cannot be merged.
Further, although they are infrequent, there is a non-negligible number of chains of size 100 and above.

\observ{Long chains, with up to 1,000 backing files, do exist. The chain size threshold triggering streaming will cap the maximum size of many chains in the infrastructure.  }

\paragraph{Chain Sharing.}
\label{sharing}

\begin{figure}
    \center
    \includegraphics[width=0.45\textwidth]{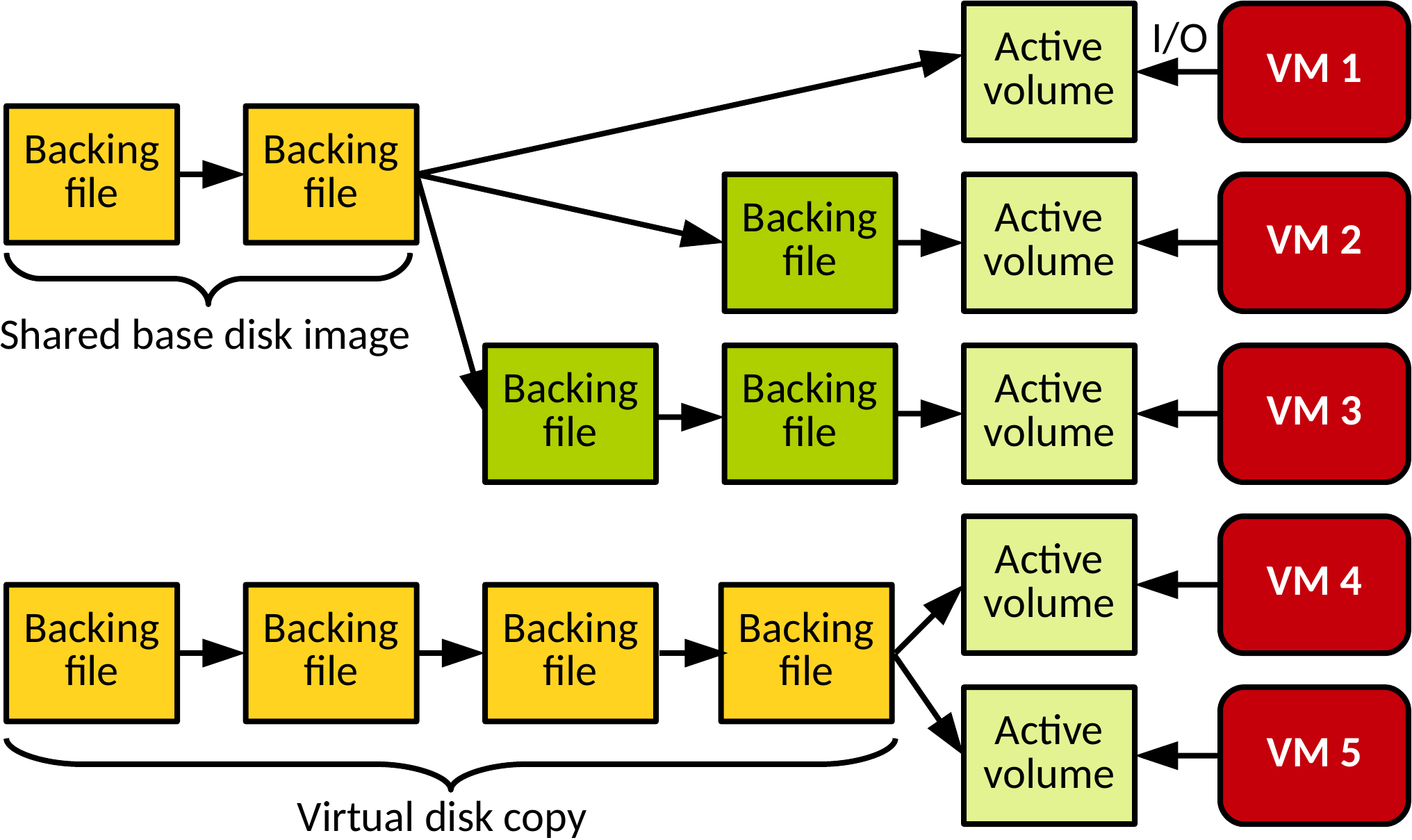}
    \caption{Chains can share backing files following a virtual disk copy, or when  using a common base image.}
    \label{fig:sharing}
\end{figure}

Certain backing files are shared, and belong to chains corresponding to different virtual disks.
The two main sources of sharing are virtual disk copy operations, as well as the use by multiple VMs of virtual disk base OS distribution images offered by the provider.
A virtual disk copy is made by transforming the active volume into a backing file, and creating 2 new active volumes on top, forming 2 chains: all the backing files are thus shared between the 2 chains.
This is illustrated on the bottom of Figure~\ref{fig:sharing}.
Concerning base OS distribution images, they are generally themselves composed of multiple snapshots corresponding to the different construction steps followed by the provider: the corresponding backing files are thus shared between all chains using a given base image.
This is illustrated on top of Figure~\ref{fig:sharing}.

\begin{figure}
    \center
    \includegraphics[width=0.45\textwidth]{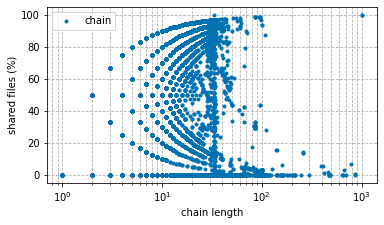}
    \caption{For each chain C of a daily measurement, percentage of backing files shared with another  chain according to the length of C.}
    \label{fig:chain-size-shared}
\end{figure}

In Figure~\ref{fig:chain-size-shared}, each point corresponds to a chain of the daily measurement previously considered.
The chain's length is indicated by its X value, and the number of backing files in the chain that are shared with at least another chain is indicated by the Y value.
Note that in theory, a chain of length $N$ can share from 0 up to $N-1$, files with other chains, i.e. all backing files without counting the active volume.
Overall, the degree of sharing is highly variable among chains.
We can observe a significant amount of chains of variable length with no sharing at all (Y value of 0).
The high number of chains with a length $N < 30$ allows us to witness, for these chains, almost all possible degrees of sharing (from 0 to $N-1$).
The large number of points around size 30 corresponds to the high number of chains of that length, due as explained above to the streaming threshold being set to 30.
Although the number of chains of size superior to 30 is smaller, one can still observe a variable degree of sharing for some of these.
Note that, base OS images are generally made of around 5 chained backing files, so most of the sharing presented in Figure~\ref{fig:sharing} is due to virtual disk copies.

\observ{Backing files can be shared between several chains when multiple VMs use the same base OS image, or to achieve virtual disk copy. The degree of sharing among chains in the infrastructure is highly variable. Past a certain length (5+), most of the sharing is due to virtual disk copies.}

\paragraph{Snapshot Creation Frequency.}
\label{snapshot-creation}

Finally, we investigated the frequency of snapshot (i.e. backing files) creation.
We looked in our daily measurement, for each snapshot creation operation, the time elapsed since the creation of the previous link in the chain (either a backing file, or the active volume for a first snapshot).

\begin{figure}
    \center
    \includegraphics[width=0.45\textwidth]{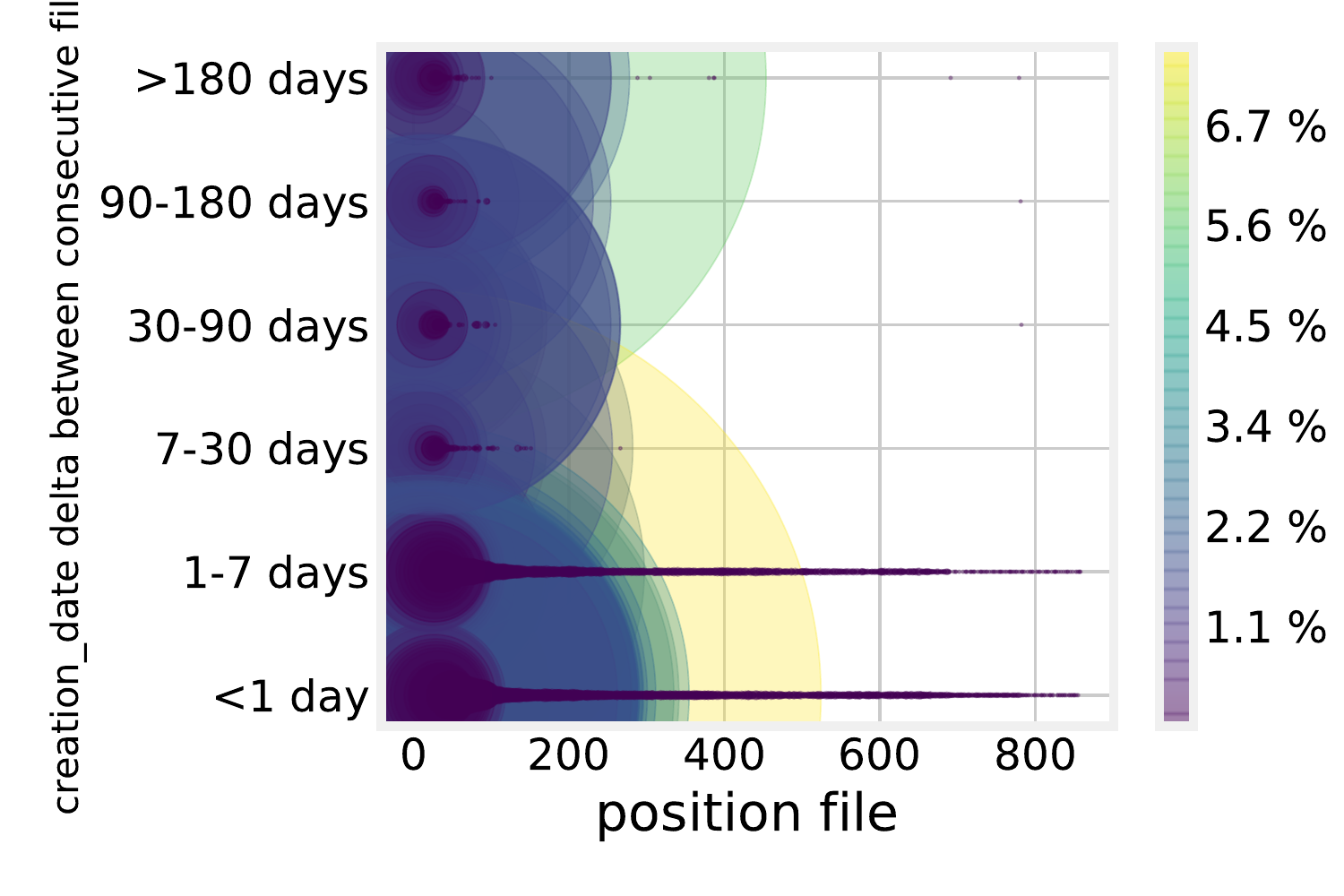}
    \caption{Snapshot creation frequency.}
    \label{fig:freq}
\end{figure}

This data is presented on Figure~\ref{fig:freq}.
Each point corresponds to a set of snapshot creation operations, placed on the Y axis into buckets corresponding to different elapsed time windows since the last link creation.
Each point's X value corresponds to the position in the chain of the created backing files.
Finally, the size and color of each point denotes how many snapshot creation operations are represented by the point, as a percentage of the total number of operations counted in our daily measurement.

As one can observe, the majority of the snapshots are made on chains of size inferior to 30.
This is due to the high number of chains of these sizes, stemming from the streaming threshold set to 30.
Further, although the frequency of snapshot creation is overall highly variable, an important number of snapshots are created with a relatively high frequency (daily or more).
Past work~\cite{OUTSCALE} noted peaks at up to 58 snapshots per hour.
One can also observe that the long chains are the result of relatively frequent (daily/weekly) snapshotting done by clients (i.e. non-mergeable through streaming).

\observ{Although the snapshot creating frequency varies widely among chains, a non-negligible amount of chains experience high frequency snapshotting.
Long chains belong to this subset, with daily/weekly snapshot created.
These snapshots are made by clients and cannot be merged with streaming.}

\section{Problem with Long Snapshot Chains}
\label{sec:problem}

\subsection{Origins of Long Snapshot Chains}
\label{cause-long-chains}
The emergence of long snapshot chains in modern virtualized environments is due to a combination of factors.
First, for data backup/fault tolerance purposes, most cloud providers offer to the client the possibility to create disk snapshots either on a regular basis, for example every 24 hours, or on-demand through an API.
The chain length will thus grow according to the snapshot frequency.
Even in the case the client deletes certain snapshots, they are kept by the provider as they form a necessary part of the Qcow2 chain backing the disk the VM in question is currently using.
Second, snapshots may be performed by the cloud provider itself due to thin provisioning strategies: virtual disk space being allocated on-demand, a disk may grow above the boundaries of the physical disk storing it and, combined with distributed storage, a snapshot allows to have the virtual disk transparently continue to grow on another physical disk without data transfer.
Although they are not visible by the client, such snapshots will be placed in the chains in the same way as the client-made snapshots and will participate to the chain's size increase.

An intuitive way to tackle the long chains issue is to merge snapshots that have been deleted and as such compact the chain, in a process referred to as streaming.
These techniques are quite limited as the cloud provider has no control over the client-made snapshots.
Furthermore, streaming seriously impacts guest I/O performance: we measured the disk latency from the guest with \texttt{ioping} on a standard SSD (WD Blue) and noted a 100x increase during streaming.
Streaming can be quite long according to the size of the merged snapshots, and a streaming operation needs to abort in case the client decides to reboot/halt the VM.

\observ{Long chains are due to the client- as well as provider-made snapshots, and to the limitations of the methods (e.g., streaming) to reduce their length.}


\subsection{Problem Statement}
\label{problem-statement}
From the illustration presented in Figure~\ref{fig:cache-walk}, the reader can intuitively see the two scalability issues posed by Qcow2 for long chains.
The first one is memory footprint increase, caused by L2 entry duplication in indexing caches.
In Figure~\ref{fig:cache-walk}, cluster 1 and cluster 2 are present in the two indexing caches.
The second consequence is the negative impact on IO request latency.
We can formalize the average cache miss cost ($Y$) using this equation:
\begin{equation}
\begin{split}
Y=[(Hit_\% \times T_M)+(Miss_\% \times (T_D+T_L+T_F))+\\(UnAl_\% \times T_F)] \times N
\end{split}
\end{equation}
where $T_M$ is the RAM access time (about 100ns), $T_D$ is the disk access time (about 80$\mu$s), $T_L$ is the time to traverse all software and network layers (about 1$\mu$s), N is the chain length, $Hit_\%$, $Miss_\%$, and $UnAl_\%$ are respectively the hit, miss and unallocated events ratios.
According to the fact that $T_D$, $T_L$ and $F_F$ are too high compared to $T_M$, even a small miss and unallocated ratio will lead to significant performance degradation~\cite{FVD}.
This degradation is exacerbated for long snapshot chains.



\begin{figure}
    \centering
    \includegraphics[width=0.4\textwidth]{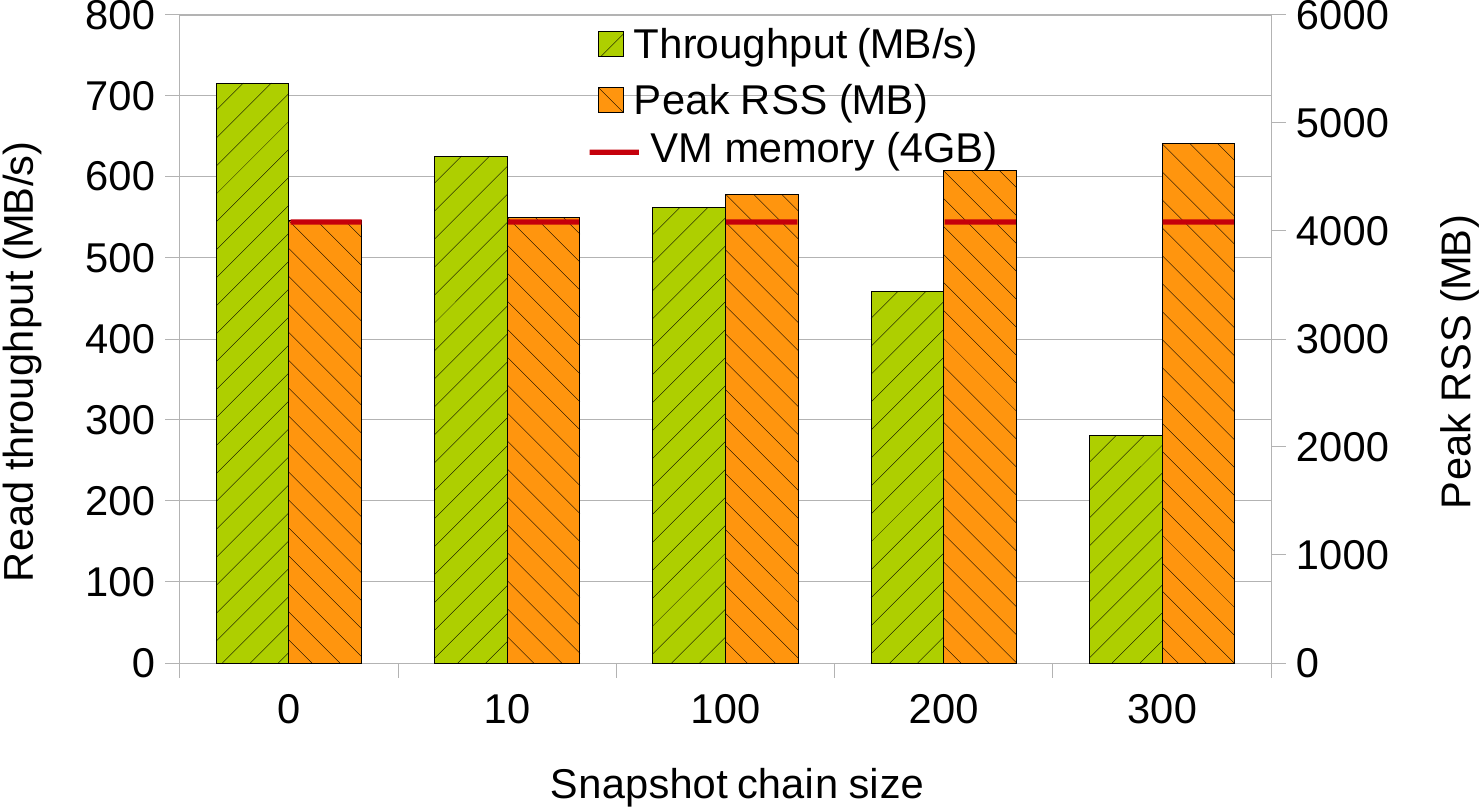}
    \caption{I/O performance and memory footprint evolution with snapshot
    chain size.}
    \label{fig:motivation}
\end{figure}

\subsection{Assessment}
\label{assessment}
A VM running on a long Qcow2 snapshot chain sees its performance and memory footprint seriously impacted.
To demonstrate these points, Figure~\ref{fig:motivation} shows the evolution of these two metrics for a VM running on a virtual disk with variable chain sizes, ranging from 0 to 300 snapshots.
The total virtual disk size is 20 GB and each snapshot contains an incremental layer of
60 MB.
All files reside locally on the host's SSD.
The VM has 4 GB of allocated RAM, 4 vCPUs and runs Ubuntu 18.04.
The read throughput is measured within the VM by reading the entire disk with \texttt{dd} right after 1) a first call to \texttt{dd} on the entire disk to ensure L1/L2 caches are fully populated and 2) a guest page cache drop to assure that the Qcow2 file is accessed.
The memory footprint is measured from the host as the hypervisor's peak Resident Set Size (RSS) observed during the execution of the \texttt{dd} command.

As one can observe, although with small chains the read throughput is not substantially impacted, when the chain size grows this metric drops significantly.
On a virtual disk with a chain size of 300, the read throughput only reaches \SI{39}{\percent} of what can be achieved on a disk with no snapshots.
Regarding memory consumption, with no or a few snapshots the memory overhead that Qemu presents on top of the 4 GB used by the VM is negligible.
However, with long snapshot chains that overhead becomes significant: with 300 snapshots, 711 MB of additional RAM are consumed by Qemu.
Third, we used the \texttt{massif} heap profiler of Valgrind to investigate memory consumption during the \texttt{dd} test on the 300 snapshots-long experiments, and discovered that the memory footprint increase is due to various data structures that are allocated on a per-snapshot basis.
The main culprit for the high memory consumption with long chains is the L2 indexing
cache.
There is one Qcow2 driver instance running in the hypervisor for each Qcow2 snapshot in a chain.
Although the maximum L2 cache size defaults to 1 MB~\cite{QCOW2_CACHE}, in our experiment we set it to 2.5 MB which is enough to manage a 20 GB disk -- setting it lower seriously impacts performance.
However, because there is one cache per driver instance and one instance per snapshot, one can conclude that the cache-related memory footprint increases linearly with the number of snapshots in the chain.
These numbers were gathered on Qemu 4.2 but we also confirmed this behavior on the latest (v6.0) version.
We focus on 4.2 in the rest of this paper as it is the version used by our cloud provider partner.

We also profiled the Qemu hypervisor from the host during the execution of the aforementioned \texttt{dd} test on the 300 snapshots-long case and found that the guest only executes for \SI{7}{\percent} of the time.
Qemu's disk driver threads consume the remaining time.

\observ{Long chains lead to memory footprint and IO performance scalability issues.}

%% file: 04-contributions.tex
\section{\sys: Scalable Qemu}
\label{design}

This section presents a new version of both Qemu and Qcow2 which tackles the two scalability issues identified in the previous section, regarding performance and memory consumption.
Ideally, both metrics should be as independent as possible from the length of the backing file chain length.

\subsection{Principles and Challenges}
\label{sec:principles-challenges}
\begin{figure}
	\center
	\includegraphics[width=0.45\textwidth]{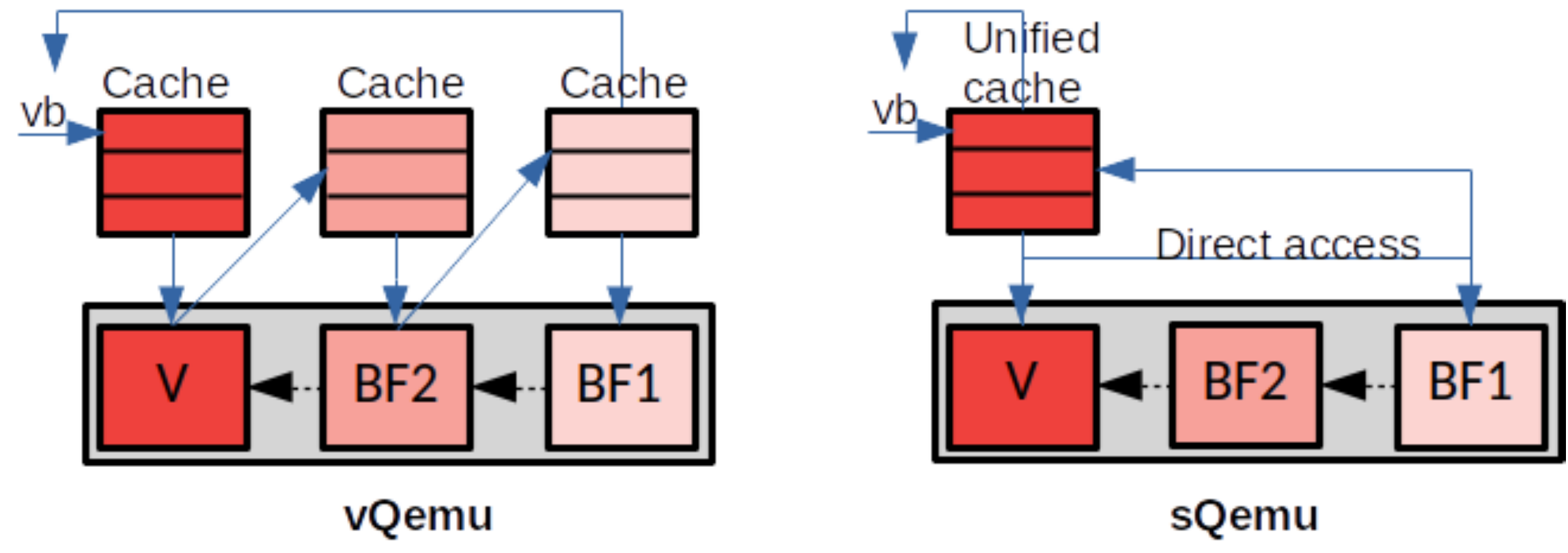}
	\caption{Qemu vanilla (left) compared to our scalable Qemu design, which follows two principles: direct access and unified indexing cache.}
	\label{fig:qemu-sqemu}
\end{figure}

\sys relies on two key principles, illustrated on Figure~\ref{fig:qemu-sqemu}:
1) direct access to on-disk indexing/data clusters, regardless of their position in the chain, upon guest I/O requests;
2) the use of a single unified indexing cache, avoiding cache entries duplication by being independent of the chain length.
In the rest of the document, we note \emph{vQemu} and \emph{vQcow2} respectively \emph{vanilla Qemu} and \emph{Qcow2} current format.
We apply the first principle through a slight but backward-compatible modification of vQcow2, requiring the storage of additional metadata in virtual disk images, as well as an update to the Qcow2 driver in the Qemu storage stack: we call that evolution \emph{\sys} for \emph{scalable Qemu}.
Meanwhile, applying the second principle only requires a carefully modification of the Qcow2 driver.

A major challenge the implementation of \sys faces regards its transparent and fast integration within the infrastructure of our cloud partner (and within cloud infrastructures in general).
Our solution should first be compatible with the different backends that can hold disk backing files in today's cloud infrastructure: these can be stored directly on the host disk but also accessed by the host through the network and served by centralized NFS servers or distributed file systems.
Hence we propose to modify a popular existing disk format rather than propose a new one~\cite{FVD}.
A related challenge is also backward compatibility: existing Qcow2 images lacking our format's metadata should still work with our updated version of Qemu (without performance/memory consumption gains on long chains), and images using our format should also work with vanilla version of Qemu that do not run our updated Qcow2 driver (once again without gains on long chains).
Alternatively, vanilla disk images can be easily converted to our format to benefit from the performance/memory footprint enhancement on long chains.

%
%

\subsection{Format Improvement}
\label{sec:format-improvement}
When a guest issues an IO request, vQemu sequentially scans the active volume and all the backing files in the chain until the proper one is found, which is not efficient.
We propose to slightly update the Qcow2 format as well as its management algorithms in Qemu in order to eliminate that chain scanning operation.
To this end, we introduce a new metadata in the format indicating, for each data cluster, the backing file that contains the latest (i.e. valid) version of the cluster.
We call this metadata the \texttt{backing\_file\_index}.
We leverage unused bits in L2 table entries to do so.
We use 16 bits to encode \texttt{backing\_file\_index} in each L2 entry.

\subsection{Unified Cache and Direct Access}
\label{sec:unified-cache-direct-access}
With direct-access, we maintain a single unified cache for the entire disk, independently of the length of the backing file chain.
Our cache has the same organization as the vanilla Qcow2 cache presented in Section~\ref{background}.
As a reminder, a cache entry corresponds to a slice and contains:
\texttt{l2\_slice\_offset} (playing the role of the tag), \texttt{ref}, \texttt{dirty}, and the L2 entries composing the slice.
As noted in the previous section, in \sys a L2 entry contains \texttt{backing\_file\_index} in addition to the default vQcow2 values.

Contrary to the vanilla version where \texttt{l2\_slice\_offset} was specific to each backing file, in our version, \texttt{l2\_slice\_offset} is related to the active volume.
In addition, one can find in the same slice, L2 entries describing data clusters belonging to distinct backing files.
Therefore, the read and write operations are performed as follows in \sys.
Let us consider $vb$ the offset of a virtual block that the guest wishes to read.
Using the same functions a vQemu, \sys computes \texttt{l2\_slice\_offset}, \texttt{l2\_slice\_index} and \texttt{l2\_index}.
If both the slice and the L2 entry exist in the unified cache and that \texttt{backing\_file\_index} contained in the L2 entry corresponds to the active volume, then there is a cache hit and the offset of the cluster data to be read is in the L2 entry.
If \texttt{backing\_file\_index} does not correspond to the active volume, this is a cache hit unallocated.
\sys locates on disk the backing file corresponding to \texttt{backing\_file\_index} and reads from it the slice at offset \texttt{l2\_slice\_offset}.
Let $s_b$ be that slice and $s_v$ be the slice currently contained in the unified cache.
\sys traverses all $s_b$ entries and updates the L2 entries in $s_v$ with the corresponding contents in $s_b$ under the following condition:
the value of \texttt{backing\_file\_index} of the L2 entry in $s_v$ is lower or equal to that of \texttt{backing\_file\_index} of the L2 entry in $s_b$.
We call ``\textit{cache correction}'' these replacement operations.
Then it sets \texttt{dirty} to 1 in $s_v$, so that the slice will be written to disk when it is evicted from the cache.
If the L2 entry does not exist in the slice there is a cache miss and the entry needs to be allocated as in vQemu.
This means that the guest is asking for a data cluster which does not yet exist on the virtual disk.
If the slice is not yet present in cache, there is a cache miss and the slice is either fetched from the active volume if it exists, or allocated if not.
These operations are similar to vQemu.

\subsection{Snapshotting}
\label{sec:snapshotting}
In vQemu, when a snapshot is created, a new Qcow2 active volume is created, with very few information (the header, the L1 table and refcounts).
We update the snapshot creation logic to copy to the newly created active volume the entire content of both L1 and L2 tables from the previous active volume, now a backing file.
The algorithm that we implement is as follows.
Let \texttt{new\_volume} be the file that will become the new active volume and \texttt{old\_volume} the old one.
We intervene at the creation of the L1 table.
Recall that it is always located at the second cluster of any Qcow2 file.
Let \texttt{new\_l1} be the new L1 table and \texttt{old\_l1} the L1 table of \texttt{old\_volume}.
After the allocation of \texttt{new\_l1}, we parse all the \texttt{old\_l1} entries.
For each entry we create the corresponding L2 table in \texttt{new\_volume}, then we set the current \texttt{new\_l1} entry with the offset of that L2 table.
Let \texttt{new\_l2} be that new L2 table and \texttt{old\_l2} be the L2 table pointed to by \texttt{old\_l1} in \texttt{old\_volume}.
Then we copy the whole content of \texttt{old\_l2} to \texttt{new\_l2}.

As a consequence, a new active volume always contains all L2 tables of the previous backing files.
The copy of L2 tables may lengthen disk snapshotting time compared to the vanilla version.
We could have implemented a copy on-demand solution, however that would mean impacting the critical path of I/O requests.
This approach would increase tail latency as it requires chain scanning to find the valid backing file.
The evaluation results show that the disturbance brought by the snapshot operation upon guest I/O performance is largely acceptable, as the total size of L2 tables is in the order of MB.
In addition, we think that VM owners are likely to accept the small price of a slight increase in snapshotting time, to benefit from an important boost in I/O performance.

%% file: 05-evaluation.tex
\section{Evaluation}
\label{evaluations}


Here we present an evaluation of \sys, aiming to answer the following three questions:
\begin{itemize}[]
	\item[$Q_1$)] Does \sys eliminate the memory footprint scalability issue of vQemu? (\S \ref{sec:memory-footprint})
	\item[$Q_2$)] Does \sys eliminate the IO performance scalability issue of vQemu? (\S \ref{sec:low-level-metrics}-\ref{sec:high-level-metrics})
	\item[$Q_3$)] To what extent \sys increases snapshotting time and disk overhead? (\S \ref{sec:overhead})
\end{itemize}

\subsection{Evaluation Setup}
\label{sec:eval-setup}

\paragraph{Methodology.}
\label{sec:eval-methodology}
We systematically compare \sys with \vanilla.
We evaluate several configurations by varying three parameters:
the chain length (1-1,000);
the virtual disk size (50GB, 150GB);
as well as the cache size (from \SI{30}{\percent} to \SI{100}{\percent} of the cache size needed to
hold the entirety of L2 entries to index a full disk, i.e. from $1.9$ MB to $6.25$ MB for a 50GB disk size, and from $5.6$ MB to $18.75$ MB for 150 GB).
For all experiments, valid clusters are uniformly distributed on the backing files of the disk's chain.
The virtual disk is populated at \SI{90}{\percent} with random data for experiments with micro-benchmarks using the Linux \texttt{dd} command, and at \SI{25}{\percent} for experiments with macro-benchmarks using the RocksDB client~\cite{rocksDB-client}.
The release of \sys includes a highly configurable chain generation script.

Otherwise indicated, the size of the L2 cache is set so that it can hold all L2 entries to index the entire disk.
All results presented in this section are an average value of 5 runs.


\paragraph{Testbed.}
To have a representative test environment, we employ 2 servers, one being the compute node running VMs, and the other the storage node holding virtual disk files.
Each server is equipped with a 32 cores Intel Xeon Gold CPU, clocked at 2.10GHz, 192 GB of RAM, Samsung MZ7KM480HMHQ0D3 SATA SSD.
They are linked with a 10Gbps Ethernet connection.
The storage node serves the virtual disk files through NFS.
Both servers run Debian 10 with Linux 4.19.0 as host OS.
All VMs run Ubuntu 18.04 with Linux 4.15.0 and are configured with 4GB of memory, 4 vCPUs.
Otherwise indicated, the virtual disk size is 50GB.

\paragraph{Metrics and Benchmarks.}
We collect two kind of metrics, \emph{high-level} and \emph{low-level} metrics.
The former are those which directly impact the end-user perceived Quality of Service.
We consider VM startup time, memory overhead, application execution time and, I/O disk throughput.
The memory overhead is the additional memory consumed by Qemu on top of the VM's allocated pseudo physical memory.
Low-level metrics represent internal costs that help explaining high-level metrics.
They are: the total number of cache misses, the total number of cache hit unallocated, and the cache lookup latency.
The lookup latency is the time taken to find the valid offset of a data cluster in the caching system.
Storage benchmarks are run in the guests.
We use microbenchmarks, including Linux \texttt{dd} (which sequentially read the entire disk from the guest i.e. \texttt{dd if=/dev/sda of=/dev/null bs=4M}) and fio~\cite{FIO}, as well as macrobenchmarks, RocksDB-YCSB~\cite{rocksDB-client} and a measurement of the VM boot time.

\input{05.1-evaluation}

\subsection{($Q_2$) Low-level Metrics}
\label{sec:low-level-metrics}
We use the same setup as in the previous section.
\input{05.2-evaluation}

\subsection{($Q_2$) High-level Metrics}
\label{sec:high-level-metrics}

\subsubsection{Micro-benchmarks}
\label{sec:micro-benchmarks}
\input{05.3-evaluation}

\subsubsection{Macro-benchmarks}
\label{sec:micro-benchmarks}
\input{05.4-evaluation}

\input{05.5-evaluation}

%% file: 05.1-evaluation.tex
\begin{figure}
	\center
	\includegraphics[width=1\columnwidth]{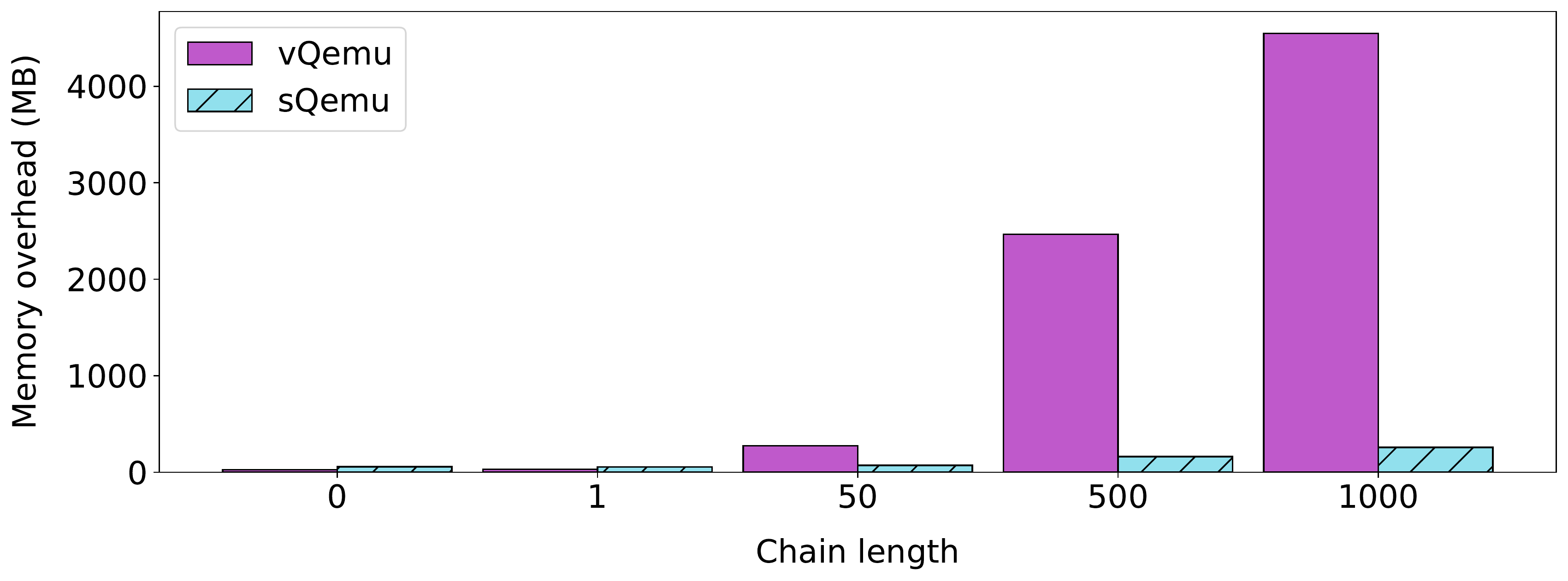}
    \caption{Memory overhead of \sys and \vanilla after reading the entire disk from the guest with \texttt{dd}, while varying chain length. \textbf{Lower is better.}}
	\label{fig:dd-memory}
\label{fig:memory-overhead}
\end{figure}

\subsection{($Q_1$) Memory overhead}
\label{sec:memory-footprint}
For this experiment we measured Qemu's resident set size after having read the entire disk from the guest using \texttt{dd}, and subtracted from this measurement the amount of RAM given to the VM (4GB) to compute Qemu's overhead.
Figure~\ref{fig:dd-memory} shows the results.
One can observe that \sys brings a significant reduction of the memory overhead when the chain length increases.
The memory savings are as follows: 205 MB for a chain length of 50 (3.9$\times$ reduction), 2303 MB for a length of 500 (15.2$\times$), and 4289 MB for a length of 1,000 (17.6$\times$).
Although it scales much better than vanilla Qemu, \sys's memory overhead still slightly increases with the chain size.
This is due to other per-snapshot data structures in Qemu that are not directly related to the caches.
Finally, note that \sys comes at the cost of a slight memory footprint increase over vanilla when the disk has no or a very small number of snapshots -- a cost that is amortized by the better scalability starting from 5 snapshots.


%% file: 05.2-evaluation.tex
\begin{figure}
	\begin{subfigure}{1\columnwidth}
	\centering
	\includegraphics[width=1\columnwidth]{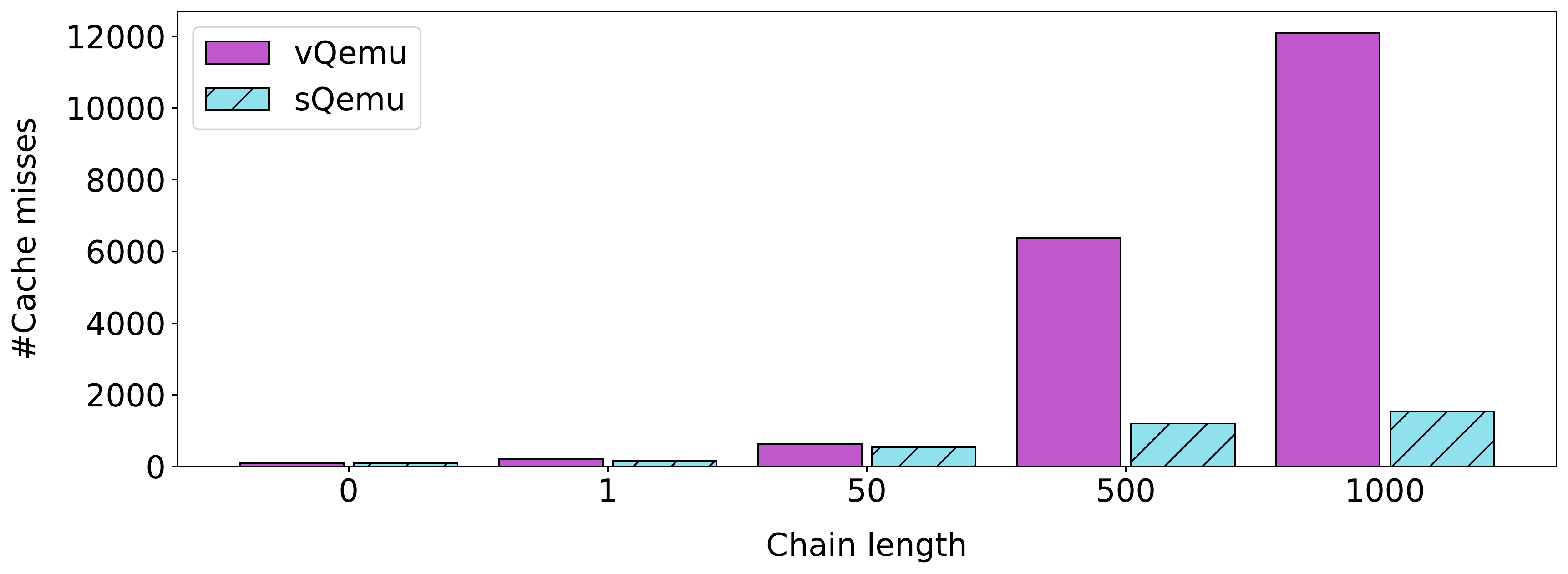}
	\caption{~}
	\label{fig:misses}
	\end{subfigure}
	\begin{subfigure}{1\columnwidth}
	\centering
	\includegraphics[width=1\columnwidth]{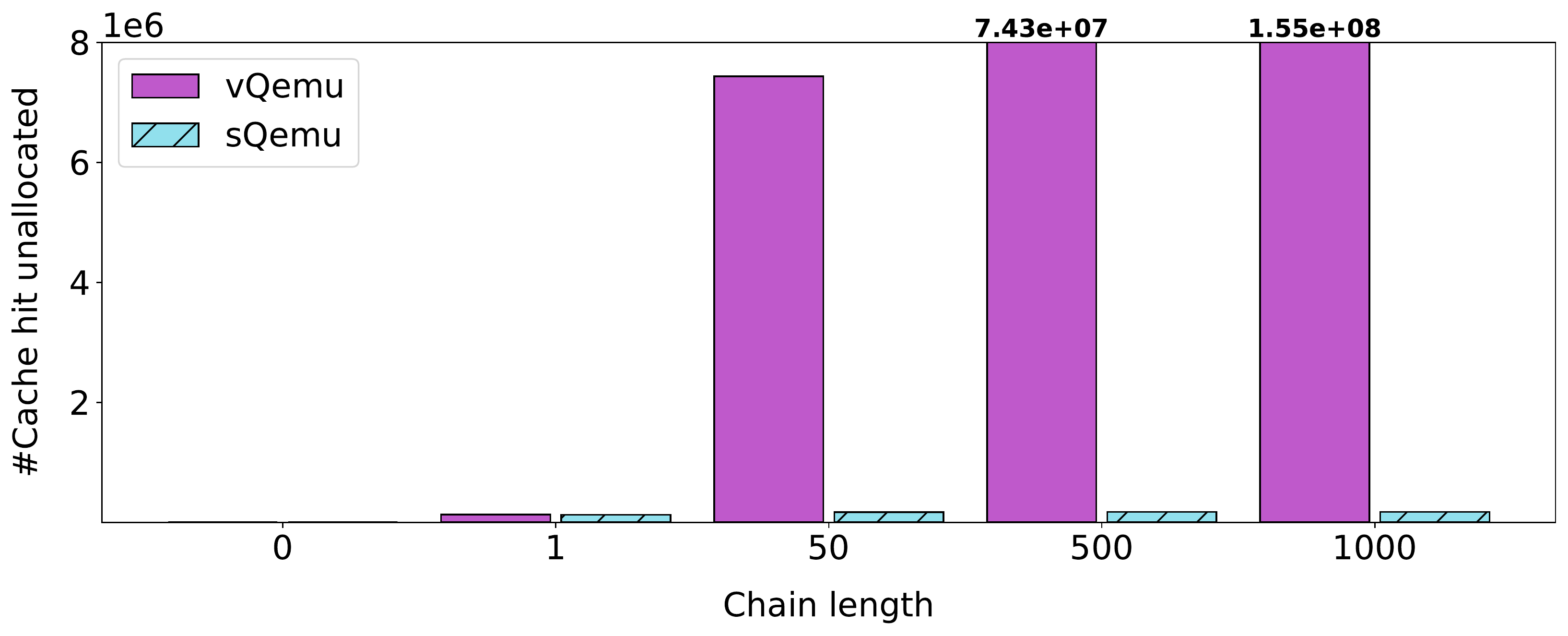}
	\caption{~}
	\label{fig:unallocated}
	\end{subfigure}
	\begin{subfigure}{1\columnwidth}
	\centering
	\includegraphics[width=1\columnwidth]{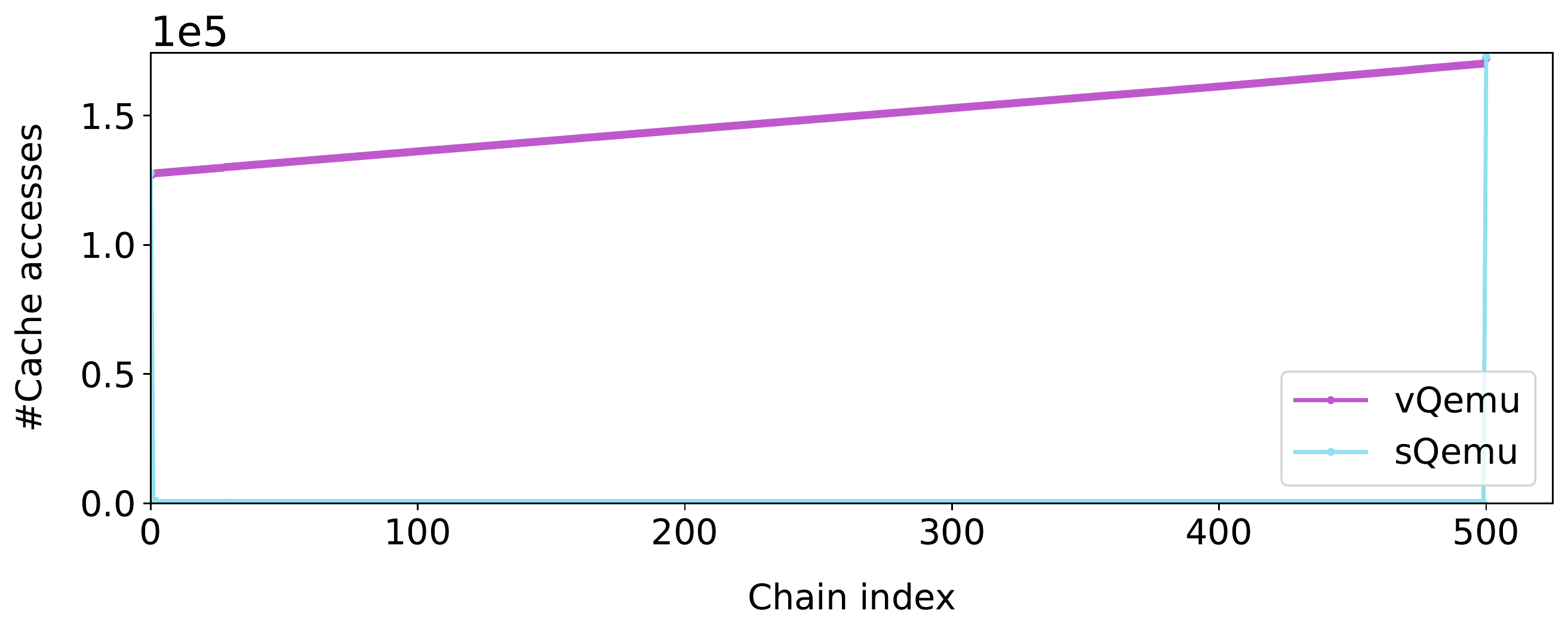}
	\caption{~}
	\label{fig:accesses}
	\end{subfigure}
    \caption{During a full disk read with \texttt{dd}, number of (a) cache misses, (b) number of cache hit unallocated,
        and (c) distribution of cache lookups according to which backing file holds the addressed data. \textbf{Lower is better.}}
	\label{fig:cache-misses-cache-hit-unallocated}
\end{figure}

\paragraph{Cache Misses and Cache Hit Unallocated.}
We instrumented \sys and vanilla Qemu to measure the number of cache misses, the number of cache hits unallocated, and the number of caches accesses per backing file of the chain.
Figure~\ref{fig:cache-misses-cache-hit-unallocated} shows the results.

We can see that \sys leads to less cache misses compared to \vanilla, see Figure~\ref{fig:misses}.
We measure up to $10 \times$ for chain length 1,000.
This difference is explained by the fact that \vanilla does not implement a cache correction mechanism as we do in \sys (see \S \ref{design}).
Therefore, when an L2 entry is only present in the cache of the backing file of index $m$ in the chain, \vanilla will generate $n-m+1$ cache misses walking the chain to get it, where $n$ is the chain length.

Concerning the number of cache hits unallocated, it is constant under \sys, see Figure~\ref{fig:unallocated}.
The increase in comparison to a virtual disk composed of a single active volume (chain length 1) is less that $1$\% for the chain length 1,000.
Concerning \vanilla, the number of cache hit unallocated increases $10,000,000 \times$ for the chain length 1,000.
This is once again explained by the fact that \vanilla looks up several caches during the chain walk.

In the experiment with a chain of length 500, we count the total number of cache lookups and plot their distribution, according to which backing file in the chain holds the requested data, on Figure~\ref{fig:accesses}.
As expected caches are much more accessed under \vanilla compared to \sys, due to the chain walks.
The gap is about 1,500\%.
The spike that appears for backing file zero, the base virtual disk image, corresponds to boot of the VM.
In fact, during that time, several IO read requests are performed on read only files (such as vmlinuz).
The spike on snapshot 500 corresponds to the accessed made on the active volume.

\paragraph{Cache Lookup Latency.}
\label{sec:low-level-metric-cache-miss}

\begin{figure}
	\centering
	\includegraphics[width=1\columnwidth]{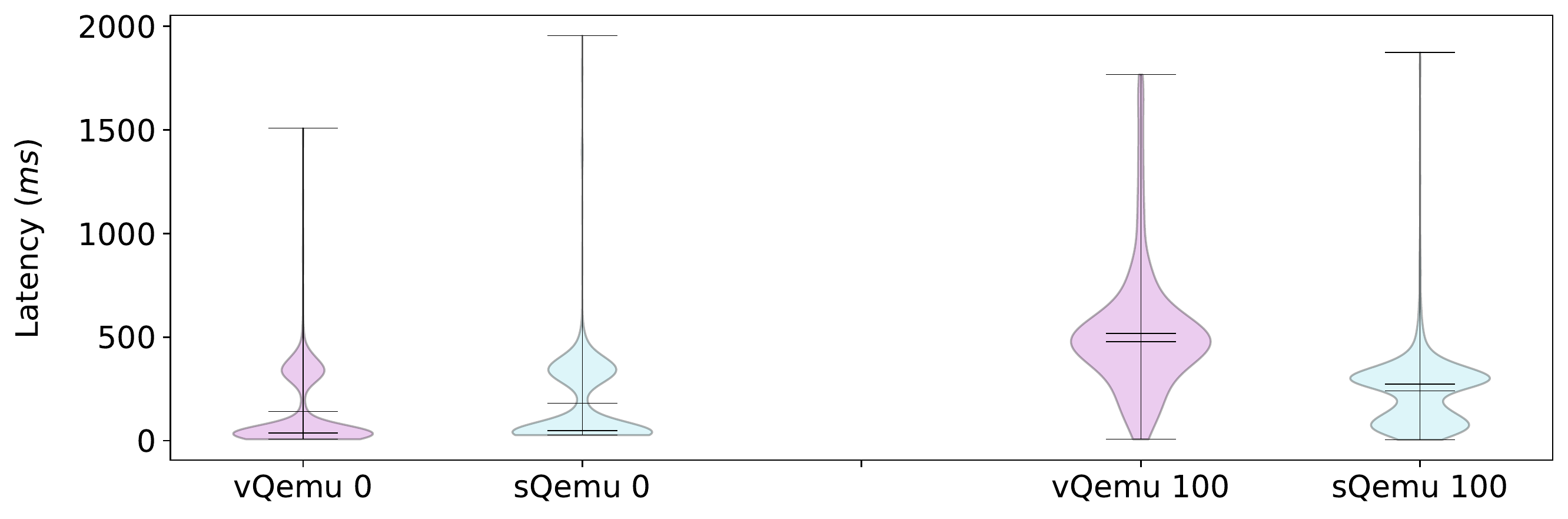}
    \caption{Cache lookup latency distribution. \textbf{Lower is better.}}
	\label{fig:cache-miss}
\end{figure}

We measured the cache lookup latency on two chain lengths: 1 and 100.
Figure~\ref{fig:cache-miss} presents the distribution of cache lookup latencies for all IO requests performed during the execution of the \texttt{dd} benchmark.
We can observe that for both systems, the mean latency value changes according to the chain length.
However, \sys leads to a better latency compared to \vanilla when the chain length increases: the mean latency is 490 ms under \vanilla and 270 ms with \sys, i.e. 1.8x faster.
Contrary to \vanilla, latency values under \sys are located around two mean values 120 ms and 270 ms.
120 ms corresponds to the cache hit mean latency while 270 ms corresponds to the cache hit unallocated mean latency.
Note that theoretically, according to the direct access principle implemented by \sys, only one value of cache hit unallocated latency can be observed compared to \vanilla.
We do not observe the same kind of distribution under \vanilla because in this experiment, data clusters are uniformly distributed over all backing files.
Therefore, most IO operations lead to a variable amount of cache hits unallocated, i.e. chain walks of variable length, according to the target data location in the chain.
This translates into highly variable and on average higher latencies in \vanilla.

%% file: 05.3-evaluation.tex
\paragraph{Disk Throughput: Linux \texttt{dd}.}


\begin{figure}
	\includegraphics[width=1\columnwidth]{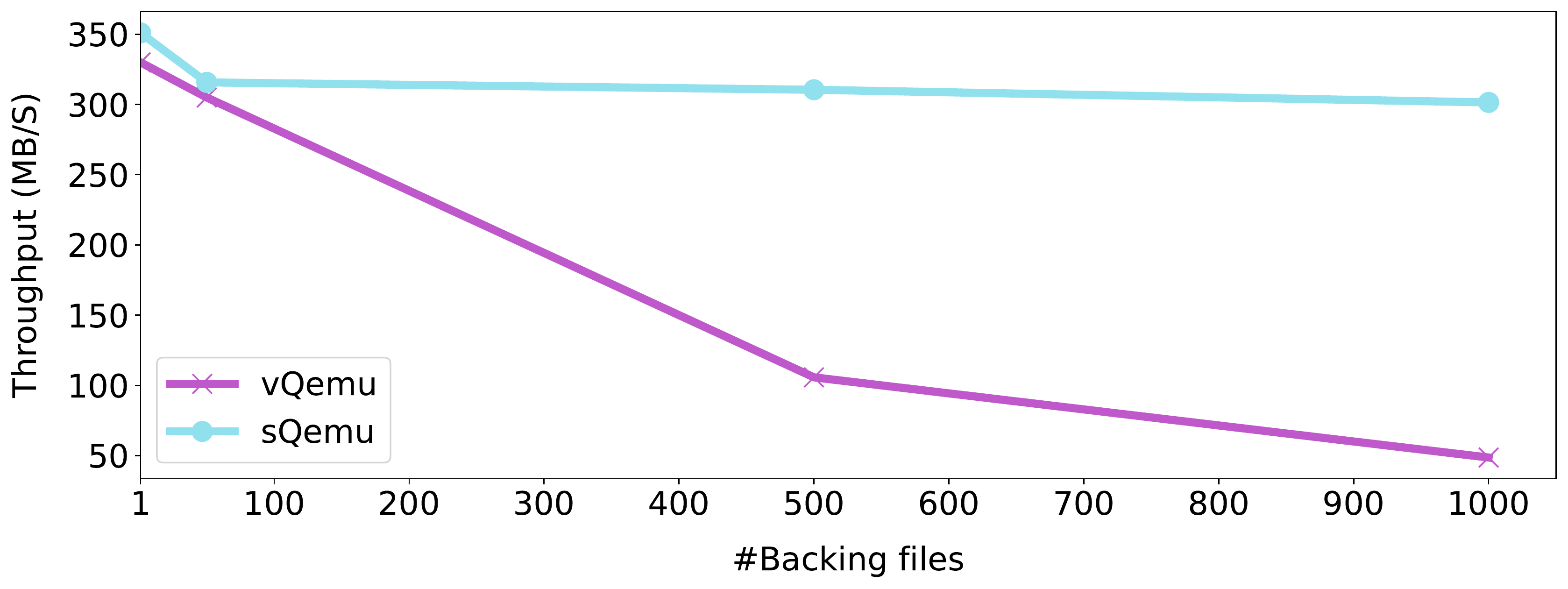}
    \caption{Throughput of Linux \texttt{dd} under \sys and \vanilla with various chain length \textbf{Higher is better.}}
	\label{fig:dd-throughput}
\end{figure}

The throughput of \texttt{dd} is presented in Figure~\ref{fig:dd-throughput} for both systems managing chains of various sizes.
We can observe no degradation under \sys while \vanilla severely degrades the throughput of \texttt{dd} when the number of backing files increases.
\vanilla incurs a slowdown of up to 84\% for the chain length 1,000.

\paragraph{Impact of the Cache Size with fio.}
We studied the effect of varying the cache size for \sys and \vanilla.
In this experiment we use of chain of length 500 and set the total cache size used by \vanilla to be equal to that used by \sys.
Because \vanilla uses one cache per layer in the chain, when \sys is given a cache size of $S$, \vanilla would get $S/L$ with $L$ being the chain length.
We vary the cache size given to each system from 1MB to 4GB, and measure the disk read throughput with fio performing random reads of small size (4 KB) on the disk node in \texttt{/dev}.

\begin{figure}
	\center
	\includegraphics[width=\columnwidth]{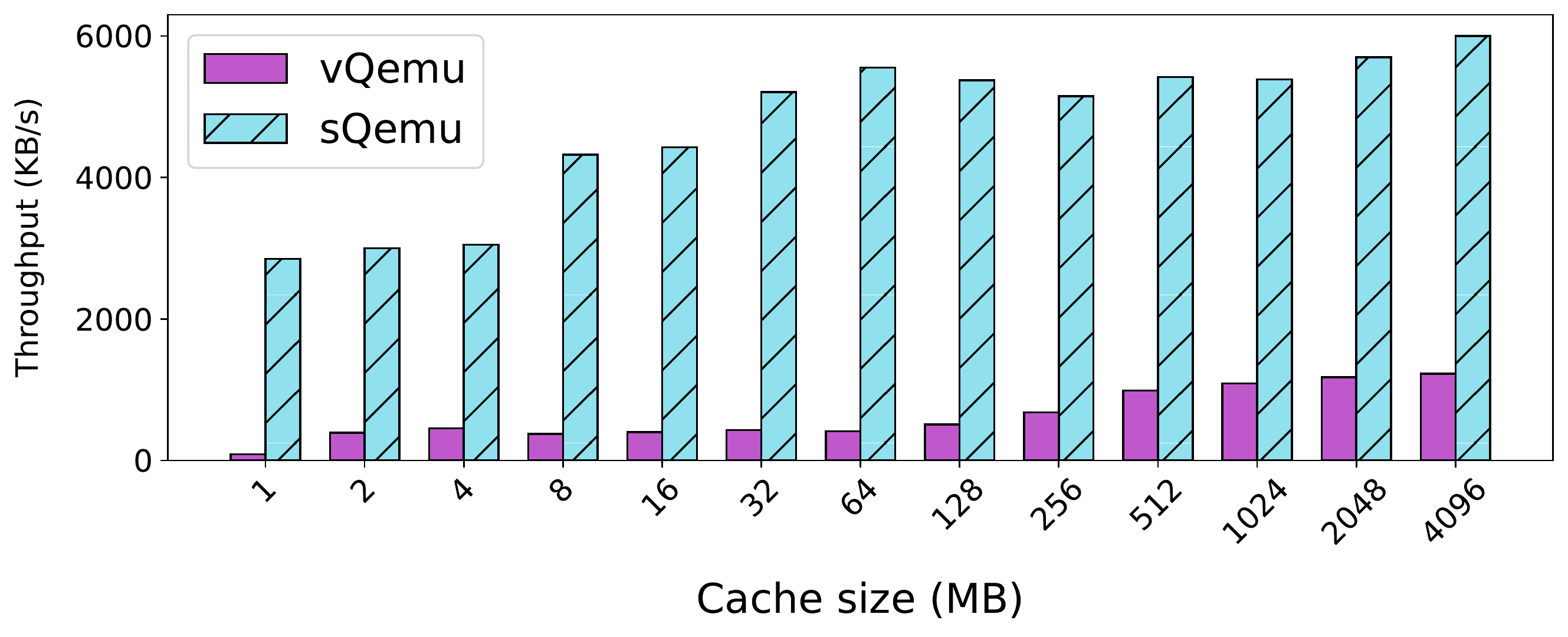}
    \caption{\texttt{fio} throughput while varying the cache size. \textbf{Higher is better.}}
	\label{fig:fio-throughput}
\end{figure}

Figure~\ref{fig:fio-throughput} shows the results.
We can observe that \sys significantly outperforms \vanilla in all cases.
With both systems, performance are sensitive to the cache size.
Concerning \vanilla, performance steadily increase up to 4 GB of cache.
This is due to the large amount of memory required by this multi-caches solution.
Regarding \sys, although peak performance are also achieved at 4 GB (6 MB/s vs. 2.5 MB/s for 1 MB of cache), from 32 MB the payback from adding more cache size diminishes significantly.
This value thus represents for a good trade-off between near-peak performance and memory footprint.
This demonstrates the high efficiency of \sys vs. \vanilla.


%% file: 05.4-evaluation.tex
\paragraph{VM Boot Time.}

\begin{figure}
	\centering
	\includegraphics[width=1\columnwidth]{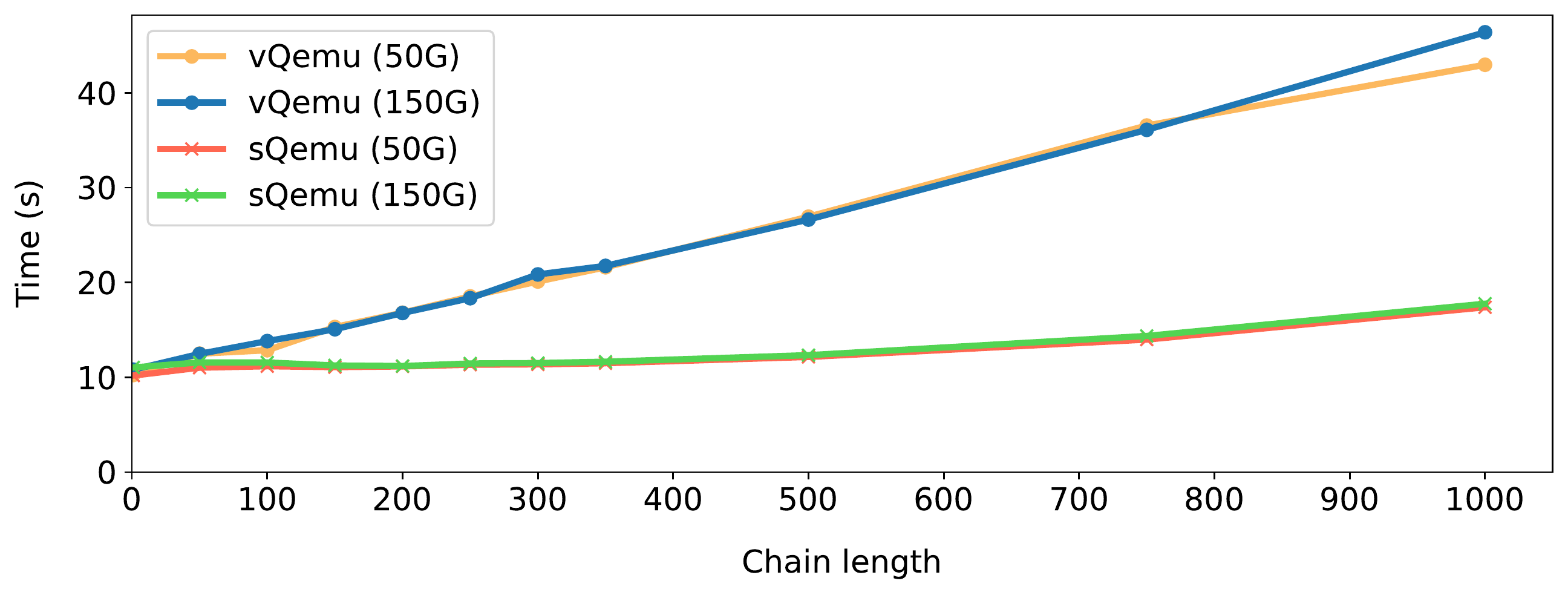}
    \caption{VM Boot Time. \textbf{Lower is better.}}
	\label{fig:boot}
\end{figure}

VM boot time is a critical metric in the cloud~\cite{MancoSOSP17, 10.1145/3050748.3050758}.
Figure~\ref{fig:boot} compares the time it takes to boot a VM under \sys and \vanilla while varying the chain length and the virtual disk size.
The boot time increases rapidly with the chain length under \vanilla: it goes from about 10 seconds on a chain of size 1 to more than 40 seconds (4$\times$) on a chain of size 1000.
On the contrary, with \sys that increase is moderate: from 10 seconds to 17 seconds (1.7$\times$).


The increase in boot time for \sys can be explained by the slight increase of the number of cache misses and cache hit unallocated discussed above.
We can see that the size of the virtual disk does not really influence the results.

\paragraph{Cloud Workload: RocksDB-YCSB.}
\label{par:high-level-rocks}

We created a RocksDB database that fills 40\% of the VM disk size, and populated using the YCSB client, generating a uniform distribution of valid clusters of the Qcow2 chains generated.
We use YCSB-C, which simulates a user performing read-only requests.
We experimented two L2 cache sizes (1 MB and 3 MB) and two chain lengths (50 and 500 snapshots).
We measured the throughput and the execution time (RocksDB's two performance metrics) of YCSB for a total of 500K requests.

\begin{figure}
	\centering
	\begin{subfigure}{0.47\columnwidth}
		\includegraphics[width=\columnwidth]{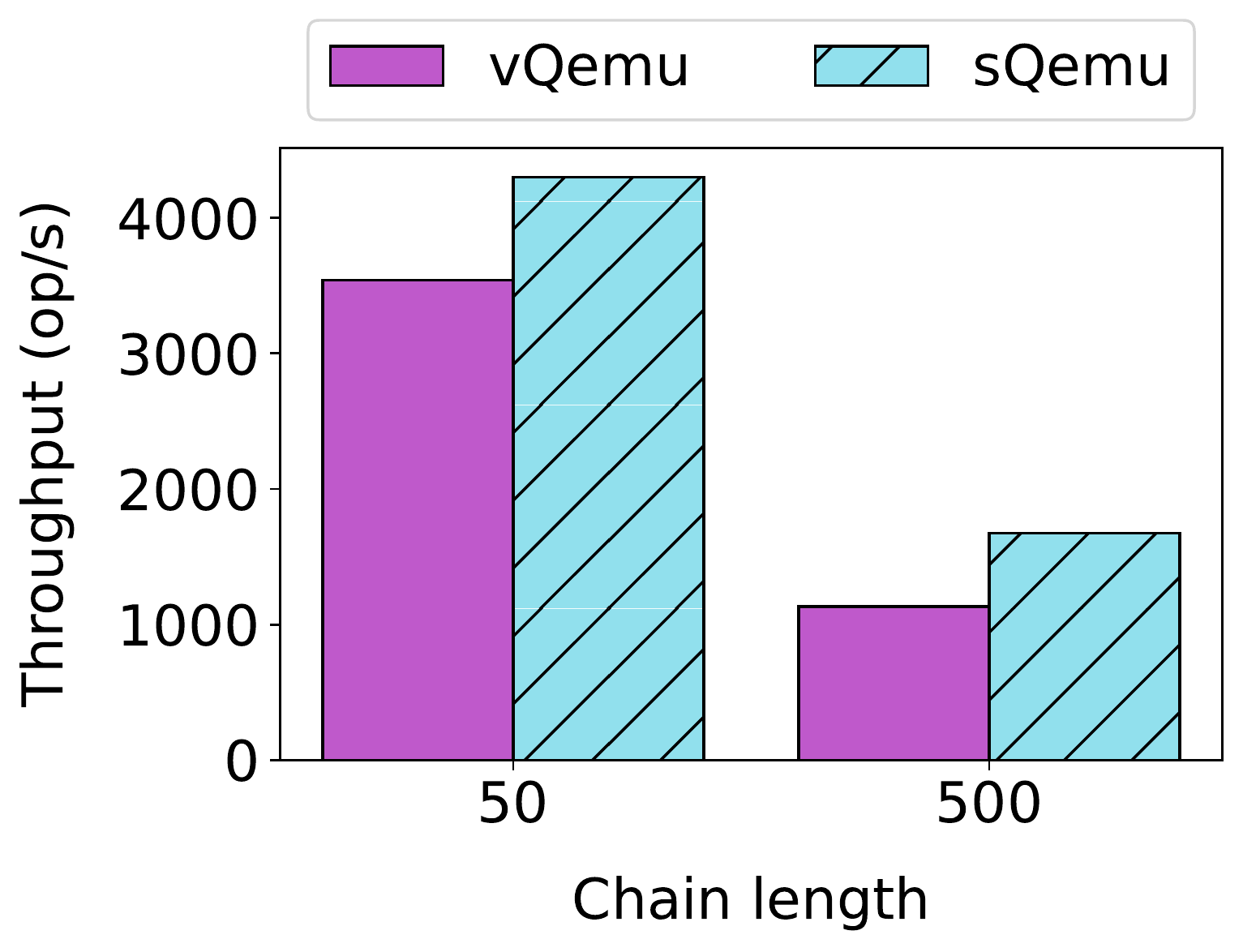}
        \caption{Throughput, with 3MB cache size. \textbf{Higher is better.}}
		\label{fig:rocks-c-throughput-cache3M}
	\end{subfigure}
	\begin{subfigure}{0.47\columnwidth}
		\includegraphics[width=\columnwidth]{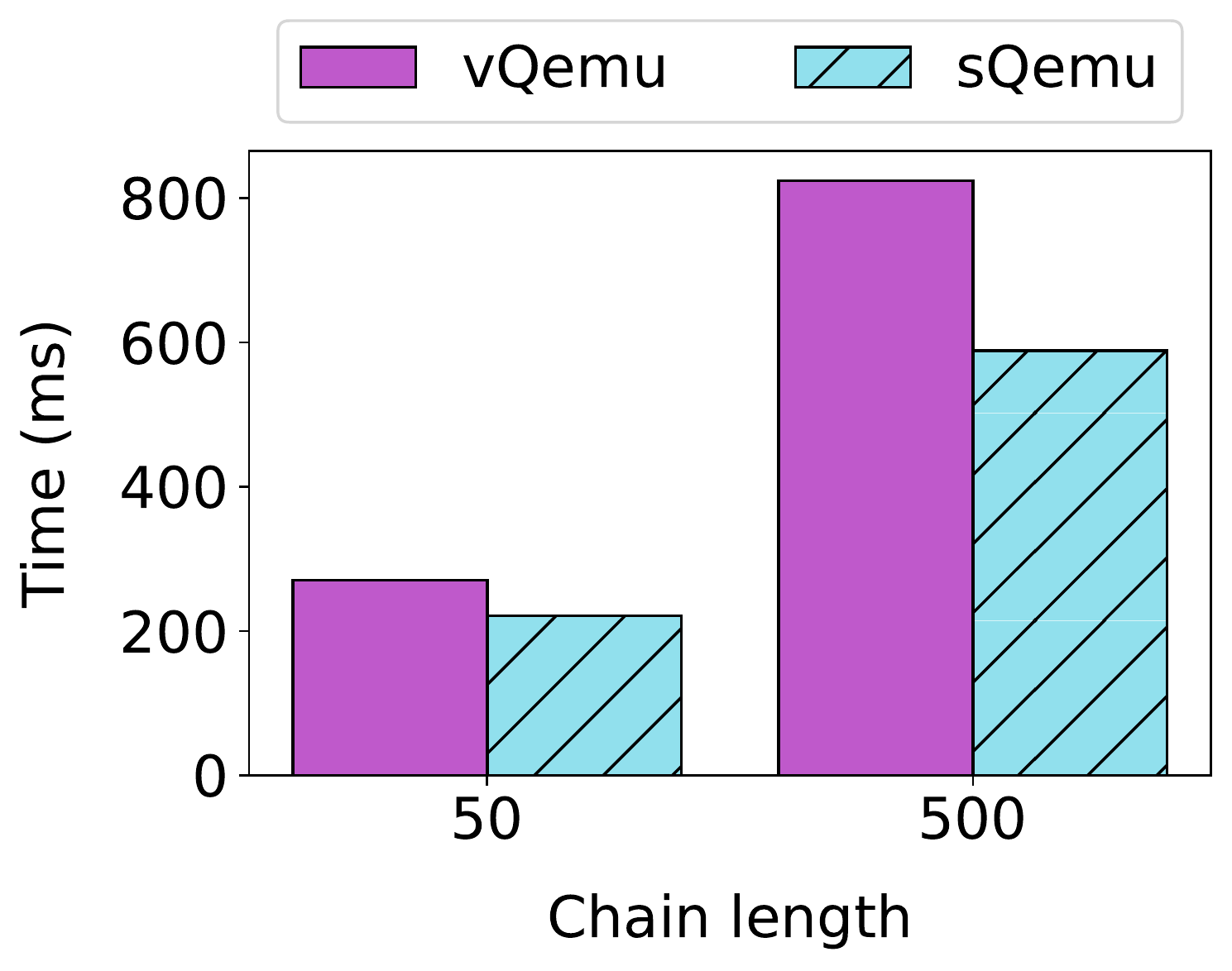}
        \caption{Execution time, with 3MB cache size. \textbf{Lower is better.}}
		\label{fig:rocks-c-time-cache3M}
	\end{subfigure}
	\begin{subfigure}{0.47\columnwidth}
		\includegraphics[width=\columnwidth]{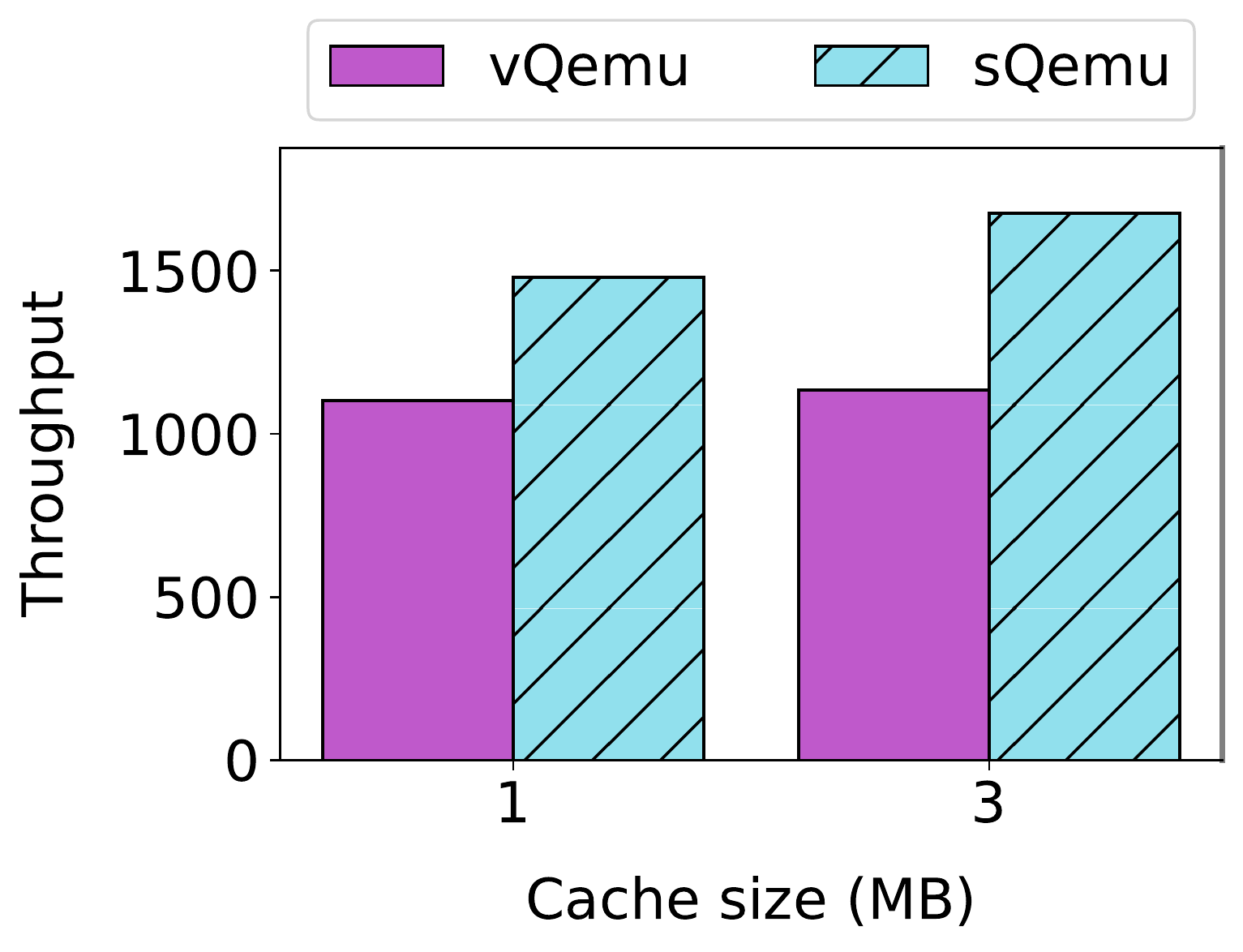}
        \caption{Throughput, with chain length 500. \textbf{Higher is better.}}
		\label{fig:rocks-c-throughput-chain500}
	\end{subfigure}
	\begin{subfigure}{0.47\columnwidth}
		\includegraphics[width=\columnwidth]{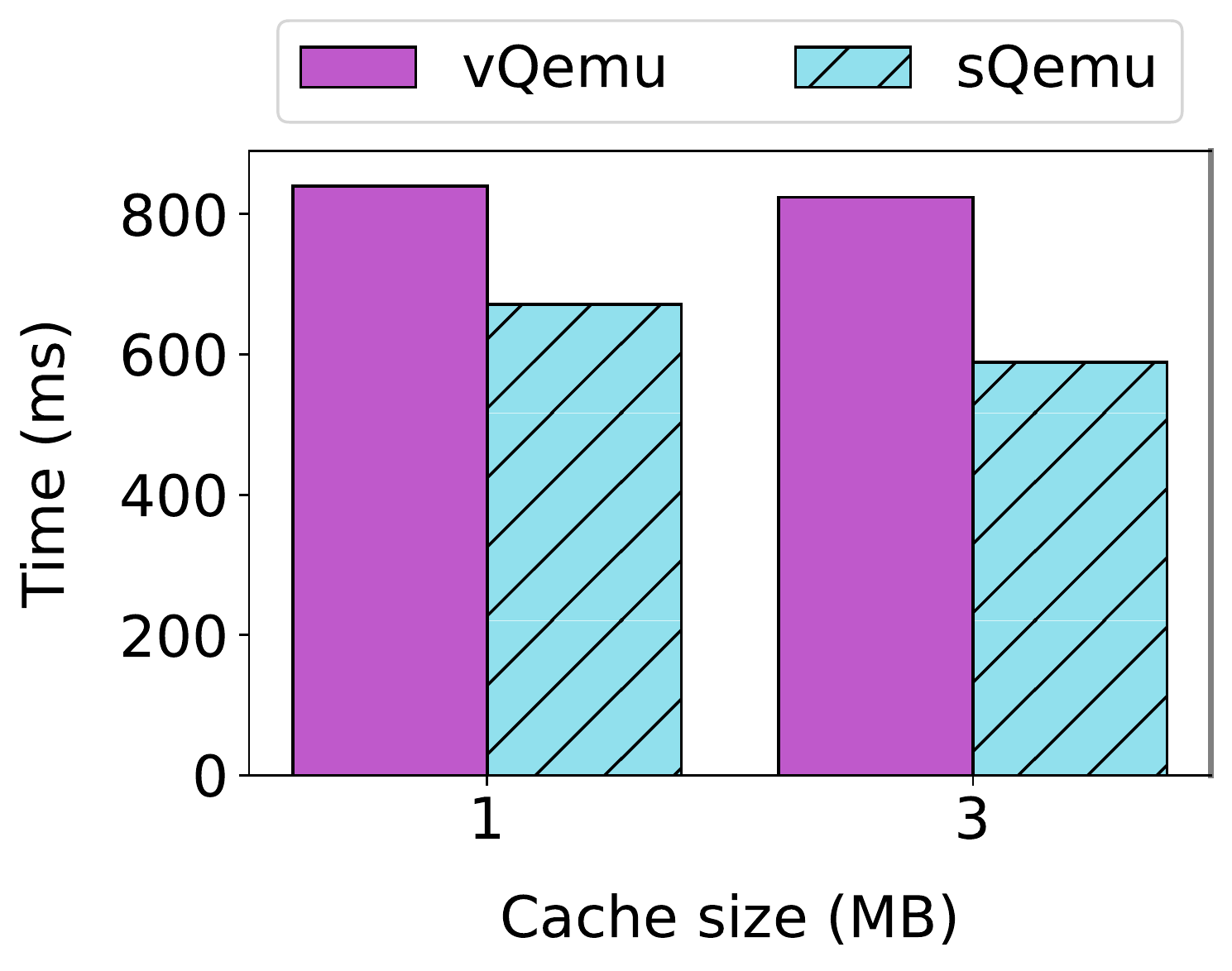}
        \caption{Execution time, with chain length 500. \textbf{Lower is better.}}
		\label{fig:rocks-c-time-chain500}
	\end{subfigure}
	\caption{RocksDB-YCSB results for YCSB-C.}
	\label{fig:rocks-c}
\end{figure}

Figures~\ref{fig:rocks-c-throughput-cache3M} and \ref{fig:rocks-c-throughput-chain500} show the results for the throughput metric.
Even if the performance of both versions decreases when the length of the chain increases, \sys still outperforms \vanilla for both chain lengths (33\% for length 50 and 47\% for length 500).
Further, with a fixed chain length at 500, the throughput of YCSB is almost constant while varying the cache size, regardless of the Qemu system.

Figures~\ref{fig:rocks-c-time-cache3M} and \ref{fig:rocks-c-time-chain500} present the execution time results.
As for the throughput, \sys improves \vanilla.
Considering a chain of 50 backing files, \sys reduces the execution time of YCSB by 36\% for 1MB cache size, and 22\% for 3MB cache size.
For a chain of 500 snapshots, the improvement is about 40\% with 1MB of cache size, and 36\% with 3MB cache size.
For throughput and execution time, the improvement of \sys over \vanilla is higher when the chain length increases.

%% file: 05.5-evaluation.tex
\subsection{($Q_3$) Overhead}
\label{sec:overhead}
As stated in \S \ref{sec:snapshotting}, when creating a snapshot under \sys, L2 tables are copied to the new created file.
This may incur two overhead types: disk usage and snapshotting time.

\paragraph{Disk space.}
The disk space overhead per snapshot depends on both VM's disk size and cluster size, as well as the
number of allocated clusters in the disk. We can model that overhead in the
worst case scenario, i.e. when every cluster is allocated (the disk is full), as follows: given
$S_{SQ}$ and $S_{VQ}$ being respectively the size of a newly created (i.e.
empty) snapshot under \sys and \vanilla, we compute the disk size of $S_{SQ}$
using the following formula:

\begin{equation}
	S_{SQ} = S_{VQ} + \frac{VM\_disk\_size}{cluster\_size} \times L2\_entry\_size
	\label{eq:formula-1}
\end{equation}

\begin{figure}
	\begin{subfigure}{1\columnwidth}
		\centering
		\includegraphics[width=1\columnwidth]{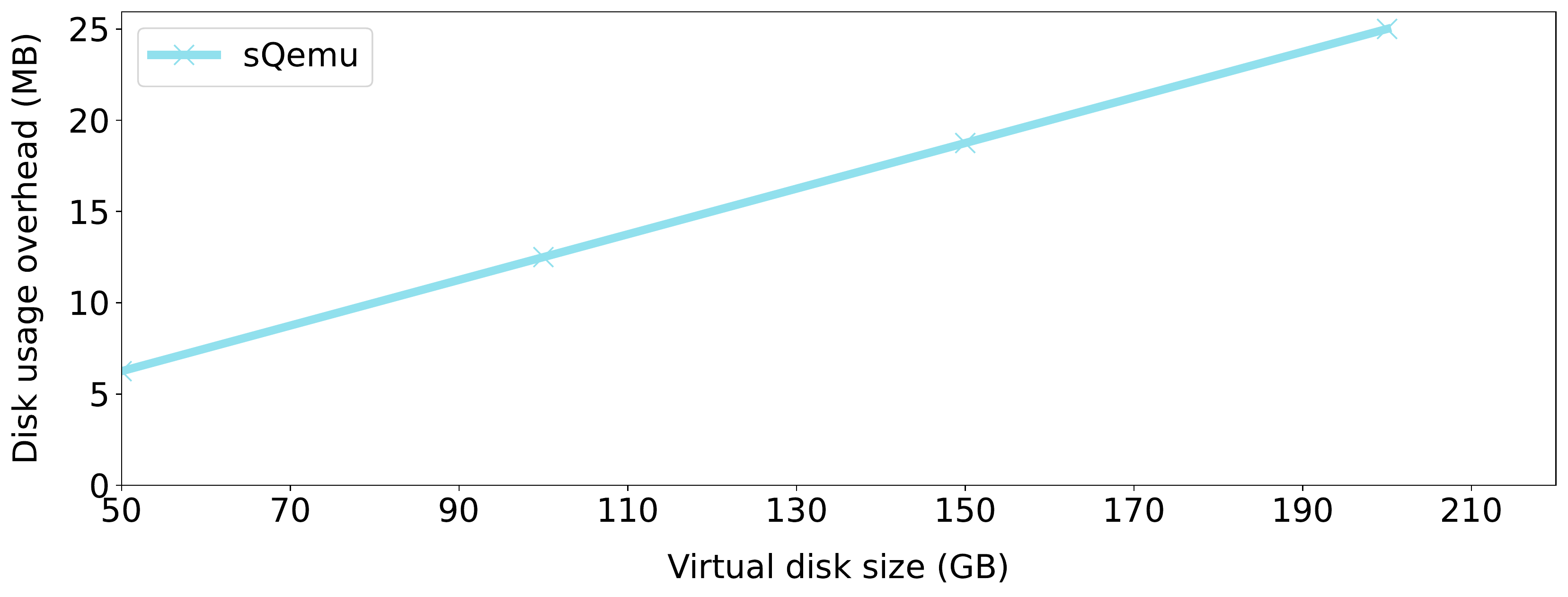}
		\caption{Disk usage overhead per snapshot.}
		\label{fig:creation_size}
	\end{subfigure}
	\begin{subfigure}{1\columnwidth}
		\centering
		\includegraphics[width=1\columnwidth]{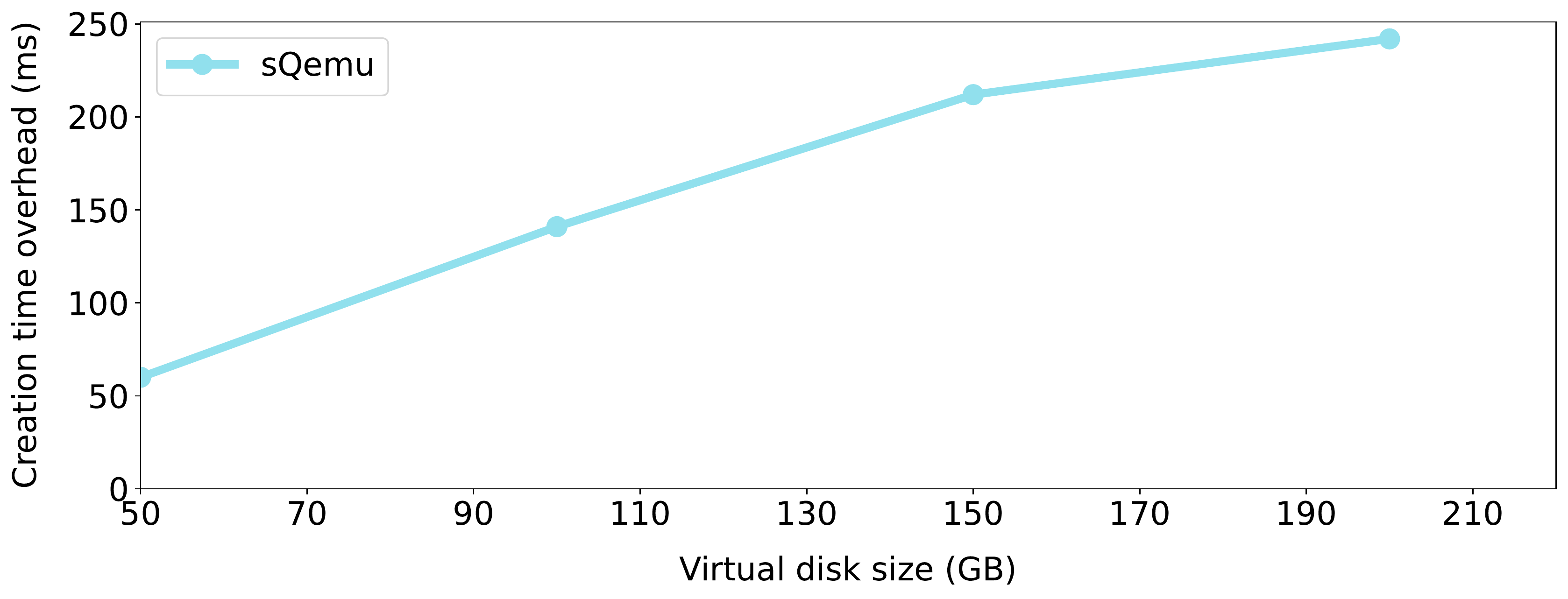}
		\caption{Snapshotting time.}
		\label{fig:creation_time}
	\end{subfigure}
    \caption{Impact of \sys on snapshotting. \textbf{Lower is better.}}
	\label{fig:sqemu-overhead-snapshot}
\end{figure}

By default, an L2 entry is 8 B, a cluster is 64 KB, and $S_{SQ}$ is 256 KB. Using
the above formula, we can compute $S_{SQ}$) while varying the VM
disk size from 50 GB to 200 GB. That per-snapshot overhead is shown in Figure~\ref{fig:creation_size}.
It increases linearly with the size of the
VM disk.
To compute the total disk overhead (still in the worst case), the per-snapshot
cost needs to be multiplied by the chain length.
Recall that from our characterization, we observed that the dominant virtual disk size in the cloud is 50 GB, giving according to our model a per-snapshot overhead is about 6 MB.
This gives a total overhead, in the worst case, of 60 MB for a chain of length 10 (0.1\% of the virtual disk size), 600 MB for length 100 (1.2\%), and 6,000 MB for length 1000 (12\%).


\paragraph{Snapshotting Time.}
We measured the time spent to create a new snapshot under \sys and \vanilla for
different VM disk sizes.
The results are presented on Figure~\ref{fig:creation_time}.
Due to the copy of all L2 entries, \sys takes much more time to create a snapshot
compared to \vanilla. For a 50GB VM, we need about 70ms to create a snapshot under
\sys and 7$\times$ less time under \vanilla. Furthermore, this overhead increases
with the VM disk size. Indeed, for a 200GB VM, the snapshot creation time under
\sys is about 12$\times$ that of \vanilla. Nonetheless, in absolute the snapshot
creation latency is quite low under \sys (in the order of ms). It allows for a relatively high
snapshot frequency.
%

%% file: 06-related-works.tex
\section{Related Work}
\label{sec:rw}

The systems software literature contains relatively few contributions regarding
storage virtualization. We identified a few recurrent topics, presented below.

\paragraph{Virtual Disk Formats.}

The research topic that is the most relevant to our work relates to proposals
of new and more efficient virtual disk formats~\cite{FVD, PARALLAX, LBS}.

Fast Virtual Disk~\cite{FVD} is a virtual disk format proposed by IBM in 2011,
that advocate for high flexibility with many configurable options and increases
I/O performance by avoiding the use of a host filesystem, reducing the size of
on-disk metadata, and using an on-disk journal. FVD supports only internal
snapshots, which means that all the chain is stored in a single file. This may
not be as flexible as the external snapshots offered by the format we focus on,
QCOW2, for example when subsets of a chain need to be stored on different
storage nodes for load-balancing or capacity reasons. It is also unclear how
FVD performs on long chains composed of hundreds or thousands of snapshots.
The system we propose is an evolution of QCOW2 which is backwards
compatible with vanilla QCOW2 disk, something that makes adoption much easier
versus proposing an entirely new format. Finally, FVD can be considered as
depreciated as it we developed for Qemu 0.14, dating from 2011, and has not
been ported to modern versions.

Parallax~\cite{PARALLAX} is a distributed architecture storing virtual disk
images that allows the use of commodity servers as storage backends, as opposed
to high-end storage arrays/switches. Among other features, Parallax offers
low-overhead and high-frequency snapshots and note, similar to our work, that
the performance overhead and memory consumption of traditional formats such as
QCOW2 increases with snapshot chain sizes. Similar to FVD, migrating an
existing cloud environment to Parallax would require significant changes to the
virtualized storage system's architecture, whereas we rely on the widely used
QCOW2 format, and are backward-compatible with environments that do not use our
system. Further, contrary to our QCOW2 format, Parallax does not support
sharing of virtual disk images, a feature heavily used in the industry to lower
storage overheads of commonly used volumes such as base images.

\paragraph{Storage Performance and Availability during VM Migration.}

Noting that VM migration significantly disrupts guest I/O performance, A few
papers~\cite{NETCHANNEL, WORKLOAD_MIG} focus on maintaining good storage
performance and availability during migration. Netchannel~\cite{NETCHANNEL}
proposes various techniques to maintain local/remote virtual disk availability
during migration. One is the ability to seamlessly switch the physical device
associated with a virtual one. Another proposed technique is the capacity for
migrated VMs initially plugged to a local disk on the host to transparently
keep using that disk through a proxy once they are migrated to another host. In
another study~\cite{WORKLOAD_MIG}, the author propose to study the storage I/O
behavior of guests to infer the most efficient data transfer schedule to reduce
disruption as much as possible during VM migration.

\paragraph{Scheduling Impact on Virtual Storage Performance.}

Several studies~\cite{CHARACTERIZATION, PARTIAL_BOOSTING} noted that VM
scheduling could have a non-negligible impact on guest I/O performance. The
authors of a study~\cite{CHARACTERIZATION} characterize the impact on processor
and I/O performance of various VM scheduler configurations, for
concurrently-running guest with CPU- and bandwidth-intensive, as well as
latency sensitive behaviors. In another paper~\cite{PARTIAL_BOOSTING}, the
authors propose a guest task-level priority boosting technique to selectively
increase the priority of I/O-bound task to increase storage performance while
maintaining CPU fairness.

\paragraph{Virtualized Storage Performance and Power Consumption.}

Other studies focus more generally on virtualized storage performance and power
consumption~\cite{POWER_CONSUMPTION, INTERRUPT_DELIVERY}. Ye and
al.~\cite{POWER_CONSUMPTION} note that existing consumption reduction reduction
techniques focusing on non-virtualized HDDs do not apply well in a virtualized
setting. They propose to bridge the semantic gap between VM and VMM through
several techniques tailored for such environments, reducing disk spin-ups and
increasing disk sleep times, in order to save energy. Another
paper~\cite{INTERRUPT_DELIVERY} focuses on the particular problem of interrupt
delivery to VMs, including the ones coming from block devices. The authors
propose an optimized interrupt delivery system for KVM. It is mostly evaluate
on network workloads but also shows moderate performance improvements on
storage workloads.

\paragraph{Cloud Storage and File Systems.}

Finally, several papers~\cite{FGCS, DEPSKY, DEPOT} focus on cloud storage and
filesystems. The Frugal Cloud File System~\cite{FGCS} proposes integrating
multiple services (AWS EBS, Azure Cache, etc.) into a single solution that aims
to be flexible from the performance and costs point of views.
DepSky~\cite{DEPSKY} introduces a cloud-based storage system targeting
security/dependability by spreading and replicating storage over multiple
clouds. Depot~\cite{DEPOT} proposes a cloud storage system that can tolerate
buggy clients and servers in order to minimize trust assumptions.

%% file: 07-conclusion.tex
\section{Conclusion}
\label{conclusion}
We present, for the first time, the characterization of virtual disk management in a large scale public cloud using the Linux-KVM-Qemu/QCOW2 virtualization stack.
Among other results, our analysis revealed the presence of long snapshot chains, leading to scalability issues for both memory footprint and performance.
We present \sys, a solution to these two issues, in the form of
a slight extension of the Qcow2 format while preserving backward compatibility.
We built \sys following the principles of direct access and single indexing cache, regardless the chain length.
We evaluate \sys extensively compare it with vanilla Qemu using a wide range of benchmarks, demonstrating that our solution effectively tackles the above issues.
For instance, \sys improves the IO throughput of RocksDB by up to 48\% compared to \vanilla, and reduce the memory footprint by 15x, when the chain length is 500.

%% file: paper.bbl
\begin{thebibliography}{10}

\bibitem{VIRT_OVERHEAD6}
Keith Adams and Ole Agesen.
\newblock A comparison of software and hardware techniques for x86
  virtualization.
\newblock {\em ACM Sigplan Notices}, 41(11):2--13, 2006.

\bibitem{harvested-vm}
Pradeep Ambati, \'{I}\~{n}igo Goiri, Felipe Frujeri, Alper Gun, Ke~Wang, Brian
  Dolan, Brian Corell, Sekhar Pasupuleti, Thomas Moscibroda, Sameh Elnikety,
  Marcus Fontoura, and Ricardo Bianchini.
\newblock {\em Providing SLOs for Resource-Harvesting VMs in Cloud Platforms}.
\newblock USENIX Association, USA, 2020.

\bibitem{XEN}
Paul Barham, Boris Dragovic, Keir Fraser, Steven Hand, Tim Harris, Alex Ho,
  Rolf Neugebauer, Ian Pratt, and Andrew Warfield.
\newblock Xen and the art of virtualization.
\newblock {\em ACM SIGOPS operating systems review}, 37(5):164--177, 2003.

\bibitem{DEPSKY}
Alysson Bessani, Miguel Correia, Bruno Quaresma, Fernando Andr{\'e}, and Paulo
  Sousa.
\newblock Depsky: dependable and secure storage in a cloud-of-clouds.
\newblock {\em Acm transactions on storage (tos)}, 9(4):1--33, 2013.

\bibitem{ressource-central}
Eli Cortez, Anand Bonde, Alexandre Muzio, Mark Russinovich, Marcus Fontoura,
  and Ricardo Bianchini.
\newblock Resource central: Understanding and predicting workloads for improved
  resource management in large cloud platforms.
\newblock In {\em Proceedings of the 26th Symposium on Operating Systems
  Principles}, SOSP '17, page 153–167, New York, NY, USA, 2017. Association
  for Computing Machinery.

\bibitem{SNU_NPB}
NASA Advanced Supercomputing~(NAS) Division.
\newblock Nas parallel benchmarks.
\newblock \url{https://www.nas.nasa.gov/software/npb.html}.

\bibitem{rocksDB-client}
{Facebook}.
\newblock A persistent key-value store for fast storage environments, 2012.
\newblock {\url{https://rocksdb.org/}}.

\bibitem{QCOW2_CACHE}
Alberto Garcia.
\newblock Qemu - qcow2 cache documentation, 2018.
\newblock \url{https://github.com/qemu/qemu/blob/master/docs/qcow2-cache.txt}.

\bibitem{VIRT_OVERHEAD7}
Abel Gordon, Nadav Amit, Nadav Har'El, Muli Ben-Yehuda, Alex Landau, Assaf
  Schuster, and Dan Tsafrir.
\newblock Eli: Bare-metal performance for i/o virtualization.
\newblock {\em ACM SIGPLAN Notices}, 47(4):411--422, 2012.

\bibitem{VIRT_OVERHEAD1}
Nikolaus Huber, Marcel von Quast, Michael Hauck, and Samuel Kounev.
\newblock Evaluating and modeling virtualization performance overhead for cloud
  environments.
\newblock {\em CLOSER}, 11:563--573, 2011.

\bibitem{STREAM}
Mc~Calpin John~D.
\newblock Stream: Sustainable memory bandwidth in high performance computers.
\newblock \url{http://www.cs.virginia.edu/stream/}.

\bibitem{PARTIAL_BOOSTING}
Hwanju Kim, Hyeontaek Lim, Jinkyu Jeong, Heeseung Jo, and Joonwon Lee.
\newblock Task-aware virtual machine scheduling for i/o performance.
\newblock In {\em Proceedings of the 2009 ACM SIGPLAN/SIGOPS international
  conference on Virtual execution environments}, pages 101--110, 2009.

\bibitem{NETCHANNEL}
Sanjay Kumar and Karsten Schwan.
\newblock Netchannel: a vmm-level mechanism for continuous, transparentdevice
  access during vm migration.
\newblock In {\em Proceedings of the fourth ACM SIGPLAN/SIGOPS international
  conference on Virtual execution environments}, pages 31--40, 2008.

\bibitem{VIRT_OVERHEAD2}
John~R Lange, Kevin Pedretti, Peter Dinda, Patrick~G Bridges, Chang Bae, Philip
  Soltero, and Alexander Merritt.
\newblock Minimal-overhead virtualization of a large scale supercomputer.
\newblock {\em ACM SIGPLAN Notices}, 46(7):169--180, 2011.

\bibitem{10.5555/2208461.2208469}
Duy Le, Hai Huang, and Haining Wang.
\newblock Understanding performance implications of nested file systems in a
  virtualized environment.
\newblock In {\em Proceedings of the 10th USENIX Conference on File and Storage
  Technologies}, FAST'12, page~8, USA, 2012. USENIX Association.

\bibitem{FIO}
Jay~F Lofstead, Scott Klasky, Karsten Schwan, Norbert Podhorszki, and Chen Jin.
\newblock Flexible io and integration for scientific codes through the
  adaptable io system (adios).
\newblock In {\em Proceedings of the 6th international workshop on Challenges
  of large applications in distributed environments}, pages 15--24, 2008.

\bibitem{DEPOT}
Prince Mahajan, Srinath Setty, Sangmin Lee, Allen Clement, Lorenzo Alvisi, Mike
  Dahlin, and Michael Walfish.
\newblock Depot: Cloud storage with minimal trust.
\newblock {\em ACM Transactions on Computer Systems (TOCS)}, 29(4):1--38, 2011.

\bibitem{MancoSOSP17}
Filipe Manco, Costin Lupu, Florian Schmidt, Jose Mendes, Simon Kuenzer, Sumit
  Sati, Kenichi Yasukata, Costin Raiciu, and Felipe Huici.
\newblock My vm is lighter (and safer) than your container.
\newblock In {\em Proceedings of the 26th Symposium on Operating Systems
  Principles}, SOSP '17, page 218–233, New York, NY, USA, 2017. Association
  for Computing Machinery.

\bibitem{PARALLAX}
Dutch~T Meyer, Gitika Aggarwal, Brendan Cully, Geoffrey Lefebvre, Michael~J
  Feeley, Norman~C Hutchinson, and Andrew Warfield.
\newblock Parallax: virtual disks for virtual machines.
\newblock In {\em Proceedings of the 3rd ACM SIGOPS/EuroSys European Conference
  on Computer Systems 2008}, pages 41--54, 2008.

\bibitem{VIRT_OVERHEAD5}
Roberto Morabito, Jimmy Kj{\"a}llman, and Miika Komu.
\newblock Hypervisors vs. lightweight virtualization: a performance comparison.
\newblock In {\em 2015 IEEE International Conference on Cloud Engineering},
  pages 386--393. IEEE, 2015.

\bibitem{RAVELLO}
Eyal Moscovici and Amit Abir.
\newblock How to handle globally distributed qcow2 chains.
\newblock KVM Forum, 2017.

\bibitem{10.1145/3050748.3050758}
Vlad Nitu, Pierre Olivier, Alain Tchana, Daniel Chiba, Antonio Barbalace,
  Daniel Hagimont, and Binoy Ravindran.
\newblock Swift birth and quick death: Enabling fast parallel guest boot and
  destruction in the xen hypervisor.
\newblock In {\em Proceedings of the 13th ACM SIGPLAN/SIGOPS International
  Conference on Virtual Execution Environments}, VEE '17, page 1–14, New
  York, NY, USA, 2017. Association for Computing Machinery.

\bibitem{CHARACTERIZATION}
Diego Ongaro, Alan~L Cox, and Scott Rixner.
\newblock Scheduling i/o in virtual machine monitors.
\newblock In {\em Proceedings of the fourth ACM SIGPLAN/SIGOPS international
  conference on Virtual execution environments}, pages 1--10, 2008.

\bibitem{VIRT_OVERHEAD3}
Pradeep Padala, Xiaoyun Zhu, Zhikui Wang, Sharad Singhal, Kang~G Shin, et~al.
\newblock Performance evaluation of virtualization technologies for server
  consolidation.
\newblock {\em HP Labs Tec. Report}, 137, 2007.

\bibitem{OUTSCALE}
Lo{\"\i}c Perennou, Mar Callau-Zori, Sylvain Lefebvre, and Raka Chiky.
\newblock Workload characterization for a non-hyperscale public cloud platform.
\newblock In {\em 2019 IEEE 12th International Conference on Cloud Computing
  (CLOUD)}, pages 409--413. IEEE, 2019.

\bibitem{FGCS}
Krishna~PN Puttaswamy, Thyaga Nandagopal, and Murali Kodialam.
\newblock Frugal storage for cloud file systems.
\newblock In {\em Proceedings of the 7th ACM european conference on Computer
  Systems}, pages 71--84, 2012.

\bibitem{QCOW2}
{Qemu Contributors}.
\newblock Qcow2 documentation, 2020.
\newblock
  {\url{https://github.com/qemu/qemu/blob/master/docs/interop/qcow2.txt}}.

\bibitem{NETPERF}
Jones Rick.
\newblock Hewlett-packard netperf benchmark.
\newblock \url{https://github.com/HewlettPackard/netperf}.

\bibitem{azure-functions}
Mohammad Shahrad, Rodrigo Fonseca, \'{I}\~{n}igo Goiri, Gohar Chaudhry, Paul
  Batum, Jason Cooke, Eduardo Laureano, Colby Tresness, Mark Russinovich, and
  Ricardo Bianchini.
\newblock {\em Serverless in the Wild: Characterizing and Optimizing the
  Serverless Workload at a Large Cloud Provider}.
\newblock USENIX Association, USA, 2020.

\bibitem{FVD}
Chunqiang Tang.
\newblock Fvd: A high-performance virtual machine image format for cloud.
\newblock In {\em USENIX Annual Technical Conference}, volume~2, 2011.

\bibitem{INTERRUPT_DELIVERY}
Cheng-Chun Tu, Michael Ferdman, Chao-tung Lee, and Tzi-cker Chiueh.
\newblock A comprehensive implementation and evaluation of direct interrupt
  delivery.
\newblock {\em Acm Sigplan Notices}, 50(7):1--15, 2015.

\bibitem{LBS}
Harvey Tuch, Cyprien Laplace, Kenneth~C Barr, and Bi~Wu.
\newblock Block storage virtualization with commodity secure digital cards.
\newblock {\em ACM SIGPLAN Notices}, 47(7):191--202, 2012.

\bibitem{VIRT_OVERHEAD4}
John~Paul Walters, Vipin Chaudhary, Minsuk Cha, Salvatore Guercio, and Steve
  Gallo.
\newblock A comparison of virtualization technologies for hpc.
\newblock In {\em 22nd International Conference on Advanced Information
  Networking and Applications (aina 2008)}, pages 861--868. IEEE, 2008.

\bibitem{POWER_CONSUMPTION}
Lei Ye, Gen Lu, Sushanth Kumar, Chris Gniady, and John~H Hartman.
\newblock Energy-efficient storage in virtual machine environments.
\newblock In {\em Proceedings of the 6th ACM SIGPLAN/SIGOPS international
  conference on Virtual execution environments}, pages 75--84, 2010.

\bibitem{WORKLOAD_MIG}
Jie Zheng, Tze Sing~Eugene Ng, and Kunwadee Sripanidkulchai.
\newblock Workload-aware live storage migration for clouds.
\newblock In {\em Proceedings of the 7th ACM SIGPLAN/SIGOPS international
  conference on Virtual execution environments}, pages 133--144, 2011.

\bibitem{10.1145/2597652.2597667}
Ruijin Zhou, Sankaran Sivathanu, Jinpyo Kim, Bing Tsai, and Tao Li.
\newblock An end-to-end analysis of file system features on sparse virtual
  disks.
\newblock In {\em Proceedings of the 28th ACM International Conference on
  Supercomputing}, ICS '14, page 231–240, New York, NY, USA, 2014.
  Association for Computing Machinery.

\end{thebibliography}
